\documentclass[a4paper,11pt]{article}
\usepackage{a4wide}
\usepackage{amsmath}
\usepackage{cite}
\usepackage[normalem]{ulem}
\usepackage{amssymb}
\usepackage{epsf}
\usepackage{amsfonts}
\usepackage{graphicx}
\usepackage{braket}
\usepackage{ragged2e}
\usepackage{appendix}
\numberwithin{equation}{section}
\usepackage{bbm}
\usepackage{xcolor}
\usepackage{subfigure}
\usepackage[export]{adjustbox}
\usepackage{siunitx,pslatex}
\usepackage{booktabs}
\usepackage{hyperref}
\usepackage{ulem}
\usepackage{caption}
\usepackage{tikz}
\usetikzlibrary{decorations.pathmorphing, decorations.markings,arrows.meta}
\usetikzlibrary{calc}
\hypersetup{
    colorlinks,
    linkcolor={green!50!black},
    citecolor={magenta},
    urlcolor={blue!80!black}
}

\begin{document}
	%%%%%%%%%%%%%%%%%%%%%%%%%%%%%%%
	
	%%%%%%%%%%%%%%%%%%%%%%%%%%%%
	
	\title{\bf Probing Quarkonium Diffusion in a Magnetized Quark-Gluon Plasma}
	\author{ \textbf{\textbf{Siddhi Swarupa Jena$^{a}$}\thanks{519ph2015@nitrkl.ac.in}, Arpan Bhattacharjee$^{a}$\thanks{arpanarindam@gmail.com}, David Dudal$^{b}$}\thanks{david.dudal@kuleuven.be}, \textbf{Subhash Mahapatra$^{a}$}\thanks{mahapatrasub@nitrkl.ac.in}
			\\\\\textit{{\small $^a$ Department of Physics and Astronomy, National Institute of Technology Rourkela, }}\\
			\textit{{\small Rourkela - 769008, India}}\\
			\textit{{\small $^b$ KU Leuven Campus Kortrijk--Kulak, Department of Physics, Etienne Sabbelaan 53 bus 7657,}}\\
			\textit{{\small 8500 Kortrijk, Belgium}}
			}
	
	\date{}
		\maketitle
	
	\begin{abstract}
		Motivated by the potential experimental relevance of magnetically affected heavy-quark diffusion, we consider here a five-dimensional nonlinear Einstein-Born-Infeld-dilaton model to not only holographically model the QCD thermodynamics in a magnetic background, but also to probe the charged inner structure of a heavy quarkonium. The dual model's gravitational equations of motion can be solved in analytical form via the potential reconstruction method. Using a variety of tools -- spectral functions, hydrodynamic expansions or hanging strings -- we study the anisotropic diffusion constants and heavy-quark number susceptibility, each time reporting closed form expressions. 
	\end{abstract}
	
	\section{Introduction}
	\label{sec1}
Heavy-ion collisions at particle accelerators such as the Large Hadron Collider (LHC) and the Relativistic Heavy Ion Collider (RHIC) provide a controlled environment to study matter under extreme conditions similar to those shortly after the Big Bang. These collisions produce energy densities and temperatures that exceed the QCD deconfinement threshold, leading to the formation of the quark-gluon plasma (QGP), a state of matter in which quarks and gluons are no longer confined within hadrons but instead form a strongly interacting, deconfined medium. The QGP behaves as a nearly perfect liquid with low viscosity and high opacity \cite{Heinz:2008tv,Nouicer:2015jrf,Krintiras:2023rdu} and serves as a laboratory for understanding the strong interactions of quantum chromodynamics (QCD). Among the key challenges is probing the transport properties of the QGP, such as the diffusion of heavy quarkonia \cite{Rapp:2008qc, He:2022ywp}, which offers insights into the medium's dynamic characteristics, meson flows \cite{Iwasaki:2021nrz}, or its response to external fields \cite{Jacobs:2004qv, Rischke:2003mt, Shu:2024vdv, STAR:2005gfr}.
    
 Transport coefficients in QGP are quantities that describe how various properties, such as momentum, energy, charge, \ldots, are transported within this exotic state of matter. These coefficients provide valuable insights into the QGP's dynamic behavior and its response to external influences. Some of the key transport coefficients include:
 \begin{itemize}
 \item \textbf{Diffusion coefficients:} fundamental transport coefficients that characterizes the rate at which particles diffuse or spread within the plasma. It quantifies how momentum, energy, and other conserved quantities are transported through random particle motion. 
 \item\textbf{Viscosity coefficients:} these include the shear ($\eta$) and bulk viscosity ($\zeta$), describing the resistance of the QGP medium against changes in flow velocity and volume, respectively. They influence the QGP's ability to flow collectively and its response to external forces. 
 \item \textbf{Conductivity coefficients:} key examples are the electrical ($\sigma$) and thermal conductivity ($\kappa$), determining the QGP's ability to conduct electric charge and thermal energy, respectively. They govern the transport of charge and heat within the plasma.
 \end{itemize}

Recent research has increasingly focused on how external electromagnetic fields, particularly magnetic fields, influence QCD-related phenomena \cite{Adhikari:2024bfa,Endrodi:2024cqn}. Studies indicate that, during the early stages of non-central relativistic heavy-ion collisions, a very strong magnetic field is generated, which can significantly influence the various phases of QCD \cite{Skokov:2009qp, Bzdak:2011yy, Voronyuk:2011jd, Deng:2012pc, DElia:2010abb, DElia:2021tfb, Tuchin:2013ie}. This field is estimated to reach strengths of approximately $\sim 0.3~\text{GeV}^2$. While it weakens after the collision, it remains sufficiently strong during the formation of the QGP to produce notable effects on several QGP-related properties \cite{Tuchin:2013apa, McLerran:2013hla}.

Magnetic fields play a crucial role in QCD. It has been shown to play a destructive role on the deconfinement and chiral phase transition, contributing to a phenomenon known as inverse magnetic catalysis \cite{Bali:2011qj, Bali:2012zg, Ilgenfritz:2013ara, Bruckmann:2013oba, Fukushima:2012kc, Ferreira:2014kpa, Mueller:2015fka, Bali:2013esa, Ayala:2014iba, Ayala:2014gwa}. Additionally, they influence the string tension of heavy quarks, either increasing or decreasing depending on the magnetic field's direction \cite{DElia:2021tfb, Bonati:2014ksa, Simonov:2015yka}. A significant impact on charge dynamics in QCD is also anticipated, leading to anomaly-induced effects like the chiral magnetic field and chiral vortical effects \cite{Fukushima:2008xe, Kharzeev:2007jp, Kharzeev:2015znc, Kharzeev:2010gd, STAR:2021mii }. These phenomena could have profound implications for the strong $CP$ problem and the generation of baryon asymmetry in the early universe \cite{Kharzeev:2020jxw}. Additionally, they may give rise to diverse thermodynamic phases and influence conventional charge transport mechanisms in the QGP. Furthermore, they play a crucial role in the anisotropic hydrodynamic framework used to describe QCD \cite{Das:2016cwd, Chatterjee:2018lsx, Gursoy:2014aka, Giataganas:2017koz}.

When examining transport coefficients like the diffusion constant, the magnetic field can have a significant impact, such as making the diffusion process direction-dependent. This means that particles may diffuse faster along the magnetic field lines than across them. This directional dependence is not just a theoretical prediction; it could have observable consequences in experiments as well. Accordingly, this paper aims to concentrate on the diffusion of a heavy quark, such as charm $c$ and bottom $b$, in a QGP environment. The diffusion of quarkonia is influenced by various other factors, including temperature and density, as well as the interactions between the quarkonium state and the constituents of the medium. Understanding the diffusion of heavy quarkonia is thus essential for interpreting experimental observations and probing the properties of the strongly interacting matter created in high-energy collisions, particularly in revealing any anisotropic features in QCD transport properties, which in turn might provide valuable information about the deconfinement transition and QCD phase diagram in the presence of a magnetic field. Let us refer to the review \cite{Iwasaki:2021nrz} for more details.

In general, modeling the Brownian diffusion of a charged particle with mass $m$ and charge $q$ in the presence of a background magnetic field $B$ can be achieved by employing the Langevin equation,
\begin{eqnarray}
    \Dot{p}= -\gamma p + q (\Dot{x} \times B) + R(t)\,.
\end{eqnarray}
Here $p=m \Dot{x}$ is the momentum and $\Dot{x}(t)=v(t)$ is the velocity of the Brownian particle, $\gamma$ is the friction coefficient, and $R(t)$ is the white Gaussian random force. The first term in the equation signifies the presence of a frictional force, resulting in energy dissipation experienced by the particle. The second term represents the Lorentz force acting on the particle. The third term accounts for a random force originating from the particle's interaction with a thermal bath. These forces collectively induce random thermal motion in the particle. The interaction between the Brownian particle and the fluid particles, occurring at a temperature $T$, facilitates an exchange of energy between them, ultimately leading to the establishment of thermal equilibrium. One way to conceptualize the particle's behavior is to consider it moving within the magnetic field, simultaneously losing energy to the surrounding medium while also experiencing random ``impulse kicks'' represented by the random force.

The random force $R(t)$ can be approximated as a series of independent impulses, each characterized by a random sign and magnitude, ensuring their average cancels out. Each impulse is considered an independent random event, implying that $R(t)$ is statistically independent of $R(t')$ for $t \neq t'$. This type of stochastic process is commonly referred to as white noise, for which
 \begin{equation}
    \braket{R_i(t)}=0, \hspace{10mm}  \braket{R_i(t_1) R_j(t_2)}= \kappa \delta_{ij} \delta(t_1-t_2)\,.
 \end{equation}
	Here, $\kappa$ is the Langevin coefficient. The two parameters $\gamma$ and $\kappa$ are not independent. This arises because both the frictional and random forces stem from a common source at the microscopic level, namely collisions with particles within the thermal bath, and are related through the following expression:
 \begin{equation}
     \gamma = \frac{\kappa}{2 m T}\,,
 \end{equation}
where $T$ is the temperature of the bath.

When we contemplate a constant magnetic field, the diffusion process is primarily influenced by the magnetic field's orientation along the transverse directions, leading to
 \begin{equation}
     D_\parallel = \frac{T}{m \gamma}, \hspace{10mm} D_\perp = \frac{D_\parallel}{1+ \frac{q^2 B^2}{m^2 \gamma^2}}\,,
 \end{equation}
 where $ D_\parallel$ and $D_\perp$ are diffusion constants in the parallel and perpendicular direction to the magnetic field, respectively. The diffusion coefficient in the direction parallel to the magnetic field is largely unaffected by the magnetic field because the Lorentz force does not act along the field direction. However, in the perpendicular direction, the diffusion constant is significantly smaller compared to $ D_\parallel$ \cite{ichimaru2018basic}. This reduction occurs because the magnetic field causes charged particles to gyrate around the field lines \cite{Satapathy:2022xdw}. 

 In the present paper, we examine how a background magnetic field influences the deconfinement phase transition and the diffusion of heavy quarks in QGP. Since the physics near the deconfinement transition is highly non-perturbative, we employ the gauge/gravity duality formalism to model strongly coupled QCD in a magnetic background \cite{Maldacena:1997re,Witten:1998qj,Gubser:1998bc}. There are two main approaches for constructing holographic models: top-down and bottom-up. Top-down models, such as the Sakai-Sugimoto model \cite{Sakai:2004cn, Sakai:2005yt}, Witten model \cite{Witten:1998zw}, Klebanov-Strassler model \cite{Klebanov:2000hb}, etc, originate from higher-dimensional string theory and are theoretically well-founded. These frameworks have proven successful in capturing key hadronic phenomena such as confinement, chiral symmetry breaking, and the meson spectrum. They have also been effectively applied to studies of proton structure and deep inelastic scattering \cite{Brower:2006ea, Brower:2010wf, Borsa:2023tqr, Jorrin:2022lua}. Nevertheless, these top–down constructions are not exact duals of QCD and still fail to capture certain essential aspects of the theory. For instance, the magnetized AdS background derived in \cite{DHoker:2009mmn, DHoker:2009ixq} does not support a Hawking/Page phase transition, making it unsuitable for capturing the QCD-like confinement/deconfinement transition. While this issue can be addressed by introducing a dilaton field by hand in a soft-wall model \cite{Dudal:2015wfn}, such modifications are not always consistent with all the equations of motion. Additionally, top-down models often struggle to reproduce the running coupling constant correctly, but there has been some recent progress in this direction as well \cite{Klebanov:2000nc, Erdmenger:2007cm, Klebanov:2000hb}. In contrast, bottom-up models are constructed phenomenologically to incorporate key QCD properties, even though they lack a direct string-theoretic origin. By now, several bottom-up holographic QCD models have been constructed that have successfully reproduced desirable magnetized QCD features aligning well with lattice QCD results \cite{Bohra:2019ebj, Bohra:2020qom, Dudal:2021jav, Gursoy:2017wzz, Gursoy:2016ofp, McInnes:2015kec, Arefeva:2020vae, Rougemont:2015oea, Finazzo:2016mhm}. For further discussions on holographic magnetized QCD, see \cite{Rougemont:2014efa, Fuini:2015hba, Cartwright:2019opv, Fukushima:2021got, Ballon-Bayona:2022uyy, Rodrigues:2017cha, Rodrigues:2017iqi, Arefeva:2023jjh, Arefeva:2021jpa, Jena:2022nzw, Shukla:2023pbp, Jain:2022hxl, Zhou:2022izh, Arefeva:2022avn, Deng:2021kyd, Chen:2021gop, Braga:2020hhs, Zhou:2020ssi, Ballon-Bayona:2020xtf, Zhao:2021ogc, Dudal:2018rki, Dudal:2016joz, Arefeva:2024xmg, Ammon:2020rvg, Cai:2024eqa, Kushwah:2025ymb}. Holographic QCD provides us with an approach to compute the diffusion constant in a fully non-perturbative manner, using black hole physics in the bulk. The presence of a magnetic field introduces further anisotropies, making the computation even more intricate. Here, we adopt a well-motivated bottom-up phenomenological model of magnetized AdS/QCD model to investigate the heavy quark diffusion.

 The diffusion of heavy quarks and Brownian motion in black hole physics can be studied using two distinct approaches.\footnote{For work related to heavy quark dissociation and energy loss in real QCD, see \cite{Singh:2017nfa, Singh:2020fsj}.}  The first method associates the Gubser-Klebanov-Polyakov-Witten (GKPW) prescription, where the retarded Green's function is computed within the AdS/CFT framework to analyze the system's response to perturbations, making it useful for studying transport properties and nonequilibrium effects\cite{Witten:1998qj, Gubser:1998bc, Son:2002sd }. The second approach utilizes the quantized scalar field states with the Hartle-Hawking vacuum, representing a black hole in thermal equilibrium, allowing the study of thermal fluctuations\cite{deBoer:2008gu, Dudal:2018rki}.

 The boundary Green's functions computed through the holographic action offer insights into the system's behavior, accurately predicting spectral functions, which detail the energy levels of bound states. Additionally, our investigation into the heavy quark diffusion coefficient involves analyzing a Green's function's behavior as the frequency approaches zero, illuminating how heavy quarks diffuse within the system over time. This analysis is grounded in the Kubo relation, a foundational concept in statistical mechanics that connects transport coefficients to correlation functions, such as for example 
\begin{equation}
    D= - \frac{1}{3 \chi}  \lim_{\omega\to 0} \sum_{i=1}^{3} \frac{G_{ii}^R}{\omega} \,.
    \label{diffuexp}
\end{equation}
Here, $\chi$ is the quark number susceptibility. Through linear response analysis, the above relation can be derived by considering how the system's conserved currents respond to small disturbances away from equilibrium, accounting for spatial variations within the system. More details on this can be found in \cite{Pasztor:2015rey, Laine:2016hma}.

The basic idea of the second approach is to analyze the fluctuations of the string worldsheet around a classical solution, which, in the relevant scenarios, corresponds to a straight string extending from the boundary into the bulk. The induced metric on the string worldsheet resembles a black hole geometry as the bulk spacetime contains an event horizon. It reduces the problem to studying two-dimensional quantum fields in curved spacetime. By quantizing these fluctuations, we can establish a connection between the quantum modes of the string and the motion of its endpoint at the boundary. This mapping enables the use of boundary correlation functions to probe the excitation spectrum of the worldsheet. Under the semiclassical approximation, the thermal properties of Hawking radiation can be linked to the Brownian motion of the string endpoint, leading to the derivation of the Langevin equation that governs its dynamics \cite{deBoer:2008gu}.

The effects of an external magnetic field on the heavy quark current can be incorporated in different ways. A straightforward method is to introduce a direct coupling between the heavy quark and the electromagnetic field. However, there are limitations of traditional electromagnetic field descriptions in capturing how heavy quark-bound states, or quarkonium, interact with external magnetic fields as the standard Maxwell action treats the quarkonium as a point-like particle. While quarkonium is electrically neutral overall, its internal structure, resembling an electric dipole, suggests it could still be affected by magnetic fields. To address this, a more sophisticated framework that considers both nonlinearity and internal structure, thereby allowing for a more refined description by effectively smearing out charge distributions, is called for. One such nonlinear model is the Einstein-Born-Infeld-dilaton (EBID) model. Incorporating Born-Infeld (BI)-like terms into electromagnetic field descriptions enables more accurate modeling of the quarkonium response to magnetic fields, accounting for its internal arrangement.

The magnetic field dependent diffusion constants have been computed using the Born-Infeld electrodynamics in a holographic setting   \cite{Dudal:2018rki}.\footnote{For works related to the holographic Brownian motion in different settings, including charged plasmas, rotating plasmas, non-Abelian super Yang-Mills (SYM) plasmas, and non-conformal plasmas, see \cite{Atmaja:2010uu, Sadeghi:2013lka, Atmaja:2012jg, NataAtmaja:2013jxi} for some examples.} Here, the soft wall model was implemented on the top-down magnetized AdS background of \cite{DHoker:2009mmn,DHoker:2009ixq}. Unfortunately, soft-wall models face several inconsistencies when it comes to effectively modeling QCD through holography. For example, the dilaton field is introduced into the action in an ad-hoc manner, and the area law of the Wilson loop is not always satisfied in these models. Additionally, the dilaton field in soft-wall models does not arise from a consistent gravity solution. Moreover, the Hawking/Page phase transition is also absent in the magnetized AdS backgrounds considered in soft-wall models \cite{DHoker:2009mmn,DHoker:2009ixq}. Since this transition corresponds to the confinement/deconfinement phase transition in the dual field theory, one therefore can not consistently define the deconfinement transition temperature. 

In this work, we remedy some of the issues mentioned above and investigate anisotropic magnetic field effects on the diffusion constants in a self-consistent magnetic field embedded EBID holographic model. In particular, we consider the holographic magnetized QCD model of \cite{Jena:2024cqs}, where closed-form magnetized AdS solutions of the EBID action,
which not only have a nontrivial and consistent profile for the dilaton field but also exhibit a magnetic field dependent Hawking/Page phase transition, were obtained. We investigate the holographic Brownian motion in the deconfined phase of this theory by the two approaches discussed above, and analyze the interplay of magnetic field and temperature on the quarkonia diffusion and susceptibility.

The structure of the paper is as follows. In Section~\ref{sec2}, we introduce the Einstein-Born-Infeld-dilaton gravity model and discuss the corresponding solutions. In Section~\ref{sec3}, we compute the heavy quark number susceptibility using the spectral function approach and analyze its behavior with temperature and magnetic field. Sections~\ref{sec4} and \ref{sec5} are dedicated to the study of diffusion coefficients, employing the hydrodynamic expansion and hanging string approach, respectively. Finally, in Section~\ref{sec6}, we present our key findings, and we conclude the paper with additional technical details provided in the Appendix.
%%%%%%%%%%%%%%%%%%%%%%%%%%%%%%%%
\section{The Einstein-Born-Infeld-Dilaton Model}
\label{sec2}
The Einstein-Born-Infeld-Dilaton action to study quarkonium physics is given by
\begin{align}
S_{EBI} = & -\frac{1}{16 \pi G_5} \int \mathrm{d^5}x  \biggl[\sqrt{-g} \ \biggl(R + \frac{f(\phi)}{(2 \pi \alpha')^2} \  -\frac{1}{2}\partial_{\mu}\phi \partial^{\mu}\phi -V(\phi)\biggr) \nonumber \\
& - \frac{f(\phi)}{(2 \pi \alpha')^2}\sqrt{-\det\biggl(g_{\mu \nu}+ 2 \pi \alpha' F_{\mu \nu}\biggr)}\biggr],
\label{actionEBI1}
\end{align}
where $G_5$ denotes the five-dimensional Newton's constant, $R$ represents the Ricci scalar, $F_{\mu \nu}$ stands for the field strength tensor of the $U(1)$ gauge field, and $\phi$ denotes the dilaton field. The $U(1)$ gauge field serves to introduce a constant background magnetic field $B$. The interaction between the $U(1)$ gauge field and the dilaton field is characterized by the gauge kinetic function $f(\phi)$, while $V(\phi)$ denotes the dilaton potential. Note that in the above type of bottom-up holographic models, the information of the number of colors $N_c$ is encoded in Newton's constant $G_5$. In general, $G_5 \sim N_{c}^{-2}$. For quarkonium diffusion, which requires investigation of the gauge field fluctuations, Newton's constant does not appear explicitly. Accordingly, $G_5$ or $N_c$ does not influence the overall behavior of diffusion constants in the large $N_c$ holographic limit.

The string parameter $\alpha'$ introduces a new dimensional scale and is related to the Born-Infeld parameter 
 $b$ by the relation $b=1/(2 \pi \alpha')$ \cite{Fradkin:1985qd, Tseytlin:1999dj}.\footnote{The relationship between $b$ and $\alpha'$ follows from the equivalence of the effective action of an Abelian vector field coupled to the open string with the Born-Infeld Lagrangian \cite{Fradkin:1985qd, Tseytlin:1999dj}.} Substituting this relation into Eq.~(\ref{actionEBI1}) and simplifying, one obtains the following modified form of the EBID action:
\begin{eqnarray}
 S_{EBI} =  -\frac{1}{16 \pi G_5}\int \mathrm{d^5}x \sqrt{-g} \left[R + f(\phi) b^2 \left(1-\sqrt{1+\frac{F_{\mu \nu}F^{\mu \nu}}{2 b^2}}\right) -\frac{1}{2}\partial_{\mu}\phi \partial^{\mu}\phi -V(\phi)\right].
 	\label{actionEBImodified}
 \end{eqnarray}
As the parameter $b \rightarrow \infty$, the action collapses to the Einstein-Maxwell-dilaton theory. For a nice review of Born-Infeld electrodynamics, we refer the readers to \cite{Alam2021Nov}.

The parameter $b$ can be determined using the Wilson loop, which now follows the area law in the confined phase in our EBID model, unlike earlier soft-wall models. This allows us to compute the string tension $\sigma$ at zero magnetic field and compare it with lattice QCD results. The estimated central value $\sqrt{\sigma}=0.43 \pm 0.02 ~\text{GeV}$, gives $b=0.44 ~\text{GeV}^2$. For more details about the Born-Infeld parameter, the reader is referred to \cite{Jena:2024cqs}. In the limit as the parameter $b$ approaches $\infty$, the action tends towards the Einstein-Maxwell theory asymptotically. We restrict ourselves here to a single gauge field, corresponding to a single type of quarkonium. 

By varying the action, one can derive the corresponding Einstein, Maxwell, and dilaton equations of motion. Remarkably, employing the following \textit{Ans\"atze} for the metric field $g_{MN}$, dilaton field $\phi$, and field tensors $F_{MN}$,
 \begin{eqnarray}
	& & ds^2=\frac{L^2 e^{2A(z)}}{z^2}\biggl[-g(z)dt^2 + \frac{dz^2}{g(z)} + dx_{1}^2+ e^{B^2 z^2} \biggl( dx_{2}^2 + dx_{3}^2 \biggr) \biggr]\,, \nonumber \\
	& & \phi=\phi(z), \ \  F_{\mu \nu}=B dx_{2}\wedge dx_{3}\,,
	\label{ansatze}
\end{eqnarray}
along with the following boundary conditions,
\begin{eqnarray}
&& g(0)=1, \ \  \ \ g(z_h)=0 \ \ \text{and} \ \ A(0) = 0 \,,
\label{bcs}
\end{eqnarray}
and combined with the potential reconstruction method \cite{Dudal:2017max, Bohra:2019ebj}, the corresponding Einstein, Maxwell, and dilaton equations of motion can be precisely solved in a closed-form expression solely in terms of a single form function $A(z)$.\footnote{For more details on the potential reconstruction method, see \cite{Mahapatra:2020wym, Priyadarshinee:2023cmi,Daripa:2024ksg}.} Here, $L$ is the AdS length scale, which we will set to one for simplicity. The detailed equations of motion are given in \cite{Jena:2024cqs}. The solutions are given as follows:
\begin{align}
A(z) & = -a z^2 \,, \label{asol} \\
g(z) & = 1-\frac{e^{z^2 \left(3 a-B^2\right)} \left(3 a z^2-B^2 z^2-1\right)+1}{e^{z_h^2 \left(3 a-B^2\right)} \left(3 a z_h^2-B^2 z_h^2-1\right)+1} \,, \label{gsol} \\
\phi(z) & = \frac{\left(9 a-B^2\right) \log \left(\sqrt{6 a^2-B^4} \sqrt{z^2 \left(6 a^2-B^4\right)+9 a-B^2}+6 a^2 z -B^4 z \right)}{\sqrt{6 a^2-B^4}} \nonumber\\
& \quad + z \sqrt{z^2 \left(6 a^2-B^4\right)+9 a-B^2} -\frac{\left(9 a-B^2\right) \log \left(\sqrt{9 a-B^2} \sqrt{6 a^2-B^4}\right)}{\sqrt{6 a^2-B^4}} \,, \label{phisol} \\
f(z) & = -\frac{2}{z^2} e^{2 B^2 z^2 + 2 A(z)} \sqrt{1 + \frac{B^2 z^4 e^{-2 B^2 z^2 - 4 A(z)} }{b^2 }} \nonumber \\
& \quad \times \left(-2 g(z)+ 2 B^2 z^2 g(z)+ 3 z  g(z) A'(z)+ z g'(z)\right)\,,  \label{fsol} \\
V(z) & = - b^2 \left(-1 + \sqrt{1 + \frac{B^2 e^{-2 B^2 z^2 - 4 A(z)} z^4}{b^2}} \right) f(z)+ \frac{1}{2} e^{-2A(z)}\bigl(-2 g(z) \left( 12 \right. \nonumber \\
& \quad -10 B^2 z^2  +4 B^4 z^4 \left.+ 6 z A'(z)\left(-3 +2 B^2 z^2 \right) +9 z^2 A'(z)^2 + 3 z^2 A''(z) \right) \nonumber \\
& \quad+ z \left(\left(9 - 6 B^2 z^2 - 9 z A'(z)\right)  g'(z) - z  g'(z) \right)\bigr) \ ,
\label{EBIsolutions}
\end{align}
where $z$ is the holographic radial coordinate that runs from the asymptotic boundary to the horizon radius, i.e., from $z=0$ to $z=z_h$. The solution described above corresponds to a black hole, featuring a horizon located at $z=z_h$.\par
Additionally, it is worth noting the temperature and entropy expressions corresponding to the black hole solution, as they will be valuable for our later analysis of the deconfinement transition temperature:
\begin{eqnarray}
 	& & T = \frac{z_{h}^{3} e^{-3A(z_h)-B^2 z_{h}^{2}}}{4 \pi \int_0^{z_h} \, d\xi \ \xi^3 e^{-B^2 \xi^2 -3A(\xi) } } \,,   \nonumber \\
 	& & S_{BH} = \frac{V_3 e^{3 A(z_h)+B^2 z_{h}^{2}}}{4 G_5 z_{h}^3 } \,,
 	\label{BHtemp}
 \end{eqnarray}
where $V_3$ denotes the volume of the three-dimensional spatial plane.

Here, the parameter $a$ is the only free variable in the gravity solution, meaning the solution is fully specified once $a$ is determined. Following \cite{Dudal:2017max}, we set $a=0.15 ~\text{GeV}^{2}$ by ensuring that the Hawking/Page phase transition temperature (see below for more details) to about $270~\text{MeV}$ at zero magnetic field in the pure glue sector \cite{Lucini:2003zr}. From Eq.~(\ref{phisol}), this choice of $a$ imposes a restriction on the magnetic field $B$, allowing a maximum value of about $0.6~\text{GeV}$ \cite{Bohra:2020qom}.

\begin{figure}[htb!]
	\centering
	\includegraphics[height=7cm,width=11cm]{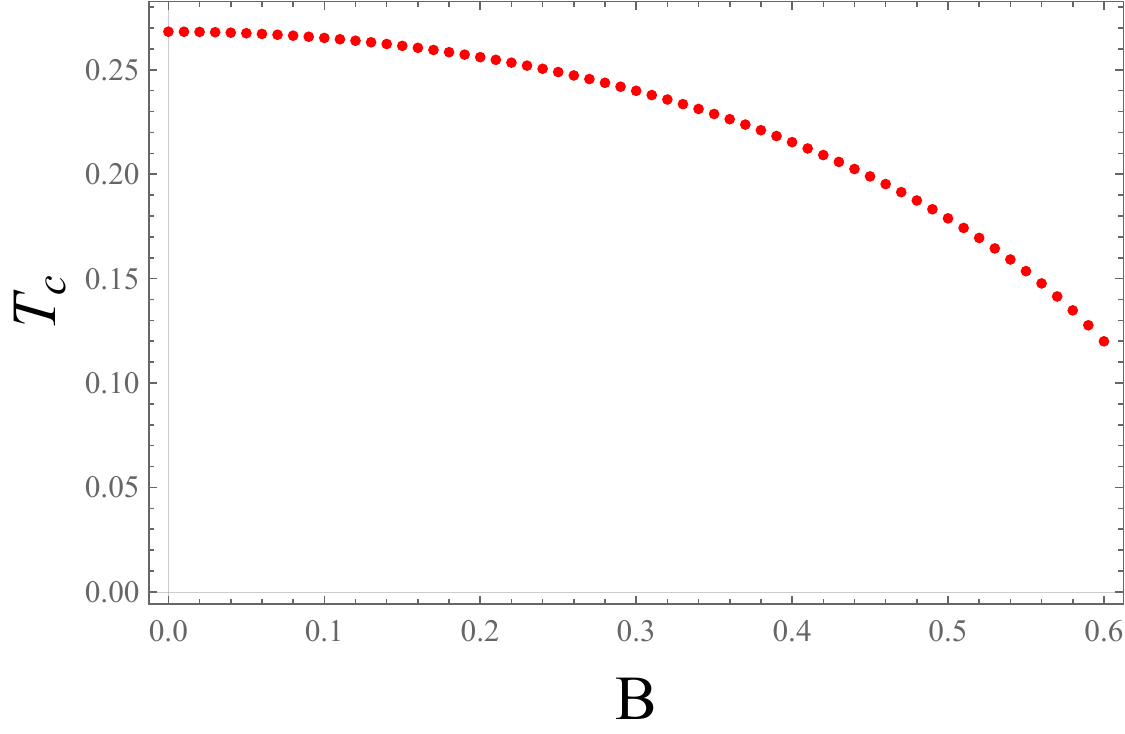}
	\caption{\small The variation of the deconfined transition temperature $T_c$ with magnetic field. In units of GeV.}
	\label{TcvsB}	
\end{figure}

In addition to the black hole solution, there is also the thermal-AdS solution, which lacks a horizon. This solution can be derived from the black hole solution by taking the limit $z_h\rightarrow\infty$. In the dual field theory, the thermal-AdS and black hole phases correspond to the confined and deconfined phases, respectively. Notably, there exists a Hawking/Page type phase transition between the thermal-AdS and black hole solutions. In particular, the thermal-AdS and black hole phases exchange dominance as the temperature is varied in the free energy landscape, with thermal-AdS dominating the phase diagram at low temperatures, while the black hole phase dominates the phase diagram at high temperatures. The corresponding transition temperature is found to be decreasing with increasing $B$. The explicit behavior of the transition temperature with $B$ is shown in Fig.~\ref{TcvsB}. The reader can check  \cite{Bohra:2019ebj} to get more background information on the black hole thermodynamics in the EMD model and on \cite{Jena:2024cqs} for EBID model. This holographic result therefore predicts a form of inverse magnetic catalysis in the deconfinement sector, qualitatively aligning with the lattice results \cite{Bali:2011qj}.

 %%%%%%%%%%%%%%%%%%%%%%%%%%%%%%
%%%%%%%%%%%%%%%%%%%%%%%%%%%%%%

\subsection{Fluctuation Equations of Motion}
To analyze the diffusion coefficients and susceptibility, we must first write the equations of motion for the relevant gauge field fluctuations in the gravity side. These equations can be derived from the Born-Infeld part of the gravity action:
 \begin{equation}
b^2 f(\phi) \left(1-\sqrt{1+\frac{F_{\mu \nu}F^{\mu \nu}}{2 b^2}}\right)\,.
	\label{BIlagrangian}
\end{equation}
This can also be rewritten as, see \cite{Jena:2024cqs},
 \begin{equation}
	\mathcal{L}_{BI} = b^2 f(\phi)\left(\sqrt{-g}-\sqrt{-\det\left(g_{\mu \nu}+ \frac{F_{\mu \nu}}{b}\right)}\right)\,.
	\label{BIlagrangianexp}
\end{equation}
In deriving the equations of motion, we split the gauge field in a classical background with a quantum fluctuation, expressed as $F_{\mu \nu}= \bar{F}_{\mu \nu} + \tilde{F}_{\mu \nu}$. Where, $\bar{F}_{\mu \nu}$ comes from the background magnetic field and it follows the same form as provided in the \textit{Ans\"atze}~(\ref{ansatze}) and $ \tilde{F}_{\mu \nu}$ is the fluctuation part. We can expand the second term of Eq.~(\ref{BIlagrangianexp}) as
    \begin{eqnarray}
\sqrt{-\det\left(g_{\mu \nu}+ \frac{F_{\mu \nu}}{b}\right)} = \sqrt{-\det\left(g_{\mu \nu} + \frac{1}{b} (\bar{F}_{\mu \nu} + \tilde{F}_{\mu \nu})\right)}  = \sqrt{-\det\left(\mathcal{G}_{\mu \nu}+\frac{\tilde{F}_{\mu \nu}}{b}\right)}\,, \nonumber\\
\approx \sqrt{-\det(\mathcal{G})}\left\{1+\frac{1}{2b}\text{Tr}(\mathcal{G}^{-1}\tilde{F}) + \frac{1}{8 b^2}(\text{Tr}(\mathcal{G}^{-1}\tilde{F}))^2-\frac{1}{4 b^2} \text{Tr}((\mathcal{G}^{-1}\tilde{F})^2)+....\right\}\,,
\label{BIlagrangianexpansion}
\end{eqnarray}
where $\mathcal{G}_{\mu \nu}=g_{\mu \nu} + \bar{F}_{\mu \nu}/b$. Explicitly, the metric tensor $\mathcal{G}_{\mu \nu}$ takes the form
\begin{equation}
\mathcal{G}_{\mu \nu} = \begin{bmatrix}
	g_{00} & 0 & 0 & 0 & 0\\
	0 & g_{zz} & 0 & 0 & 0\\
	0 & 0 & g_{11} & 0 & 0\\
	0 & 0 & 0 & g_{22} & \frac{B}{b}\\
	0 & 0 & 0 & -\frac{B}{b} & g_{g_{33}} 
\end{bmatrix} \,,
\label{metrictensor}
\end{equation}
with determinant
\begin{equation}
	\mathcal{G} = g_{00} g_{11} g_{zz} \left(g_{22} g_{33} + \frac{B^2}{b^2}\right)\,,
	\label{metricdeterminant}
\end{equation}
and inverse 
\begin{equation}
	\mathcal{G}^{\mu \nu} = \begin{bmatrix}
		\frac{1}{g_{00}} & 0 & 0 & 0 & 0\\
		0 & \frac{1}{g_{zz}} & 0 & 0 & 0\\
		0 & 0 & \frac{1}{g_{11}} &0 & 0\\
		0 & 0 & 0 &\frac{g_{33}}{X} &  -\frac{B}{b X}\\
		0 & 0 & 0 & \frac{B}{b X} & \frac{g_{22}}{X}
	\end{bmatrix}\,,
\label{metricinverse}
\end{equation}
where $X=g_{22} g_{33} + \frac{B^2}{b^2}$. 
It is convenient if we decompose the metric tensor $\mathcal{G}_{\mu \nu}$ into the symmetric ($G$) and antisymmetric ($S$) parts,
\begin{equation}
\begin{aligned}  
    G^{\mu \nu} = \begin{bmatrix}
        \frac{1}{g_{00}} & 0 & 0 & 0 & 0\\
        0 & \frac{1}{g_{zz}} & 0 & 0 & 0\\
        0 & 0 & \frac{1}{g_{11}} & 0 & 0\\
        0 & 0 & 0 &\frac{g_{33}}{X} & 0\\
        0 & 0 & 0 & 0 & \frac{g_{22}}{X}
    \end{bmatrix}\,,
    \qquad
    S^{\mu \nu} = \begin{bmatrix}
        0 & 0 & 0 & 0 & 0\\
        0 & 0 & 0 & 0 & 0\\
        0 & 0 & 0 & 0 & 0\\
        0 & 0 & 0 &0 & -\frac{B}{b X}\\
        0 & 0 & 0 & \frac{B}{b X} & 0
    \end{bmatrix}\,.
    \label{symantisym}
\end{aligned}
\end{equation}
To analyze the spectral function, we consider the vector modes $V = \frac{A_L + A_R}{2}$ and adopt the gauge choice $A_z = 0$ \cite{Dudal:2015kza}. The modes are then expressed through a Fourier expansion of the form $\sim e^{i\textbf{k} \cdot \textbf{x} -i \omega t}$.  We can obtain the fluctuation equations of motion from the gauge field equation, 
\begin{eqnarray}
\partial_{\mu}\biggl[\sqrt{-g}\frac{f(\phi)}{\sqrt{1+\frac{(\bar{F}_{\mu \nu} + \tilde{F}_{\mu \nu})(\bar{F}^{\mu \nu} + \tilde{F}^{\mu \nu})}{2b^2}}} (\bar{F}^{\mu \nu} + \tilde{F}^{\mu \nu})\biggr]=0\,.
\label{fluctuationeom}
\end{eqnarray}
At the linear order, the fluctuation equation of motion is described as,
\begin{equation}
	\partial_z^2 V_i + \partial_z \left(\ln\left(\sqrt{-\mathcal{G}} f(z) G^{zz} G^{ii}\right)\right)\partial_z V_i - \frac{G^{tt}}{G^{zz}} \omega^2 V_i = 0\,,
 \label{parallelfluceq}
\end{equation}
where $i=1$ corresponds to the fluctuation in the parallel direction, whereas $i=2$ (or $3$) corresponds to the fluctuation in the perpendicular direction to the background magnetic field.
Interestingly, as in \cite{Dudal:2014jfa}, the same equation of motion for $V_1$ and $V_2$ can be obtained from the following modified gauge equation,
\begin{equation}
\partial_{\mu}\left(f(\phi)\sqrt{-\mathcal{G}} \tilde{F}^{\mu \nu}\right)=0\,.
\label{fluctuationEOM}
\end{equation}
Using these expressions, it becomes evident that the term $\text{Tr}(\mathcal{G}^{-1}\tilde{F})$ in Eq.~(\ref{BIlagrangianexpansion}) vanishes due to the equation of motion of the background gauge field. This approach allows one to demonstrate that (\ref{parallelfluceq}) and (\ref{fluctuationEOM}) lead to the same equations of motion for fluctuations.
%%%%%%%%%%%%%%%%%%%%%%%%%%%%%%%%%%%%%%%%

 \section{Probing Quark Number Susceptibility via Spectral Function Analysis}
 \label{sec3}
	\subsection{ A Short Review}
To investigate the high-temperature phase of QCD and its phase diagram, quark number susceptibility \( \chi \) has emerged as a key observable due to its significant variation near the critical temperature \( T_c \). In addition to $\chi$, the ratio of shear viscosity to entropy density $\left(\eta/s\right)$ serves as another key probe for investigating the QGP phase diagram \cite{Lacey:2006bc, Csernai:2006zz}.  
Quarks are always coupled to gluons at low temperatures due to confinement, which modifies their properties. The dressed quark propagator accounts for these interactions and, as a result, differs from the free quark propagator. It is generally expressed as:  
\begin{equation}
    S(p) = \frac{Z(p)}{i\not{p} - M(p)}\,,
\end{equation}
where $M(p)$ represents the momentum-dependent quark mass function, arising from interactions with gluons, while $Z(p)$ is the wavefunction renormalization factor, which influences the quark's propagation probability. In non-perturbative QCD approaches, such as those based on Dyson-Schwinger equations \cite{Fischer:2014mda,Alkofer:2000wg,Maris:2003vk,Bashir:2012fs}, both $M(p)$ and $Z(p)$ are computed to investigate confinement and chiral symmetry breaking in the light flavour sector.

Mathematically, $\chi$ can be expressed as an integral involving the dressed quark propagator and the quark-gluon vertex \cite{He:2008yr, He:2008zzb}, where the vertex describes the interaction of quarks with an external vector current.  
Lattice QCD simulations have played a critical role in studying $\chi$ and its associated fluctuations in the vector channel at high temperatures. These investigations have revealed a strong correlation between $\chi$ and vector channel fluctuations, which are significantly suppressed in the QGP phase compared to the hadronic phase \cite{Kunihiro:1991qu}. This suppression is attributed to the weakening of quark-antiquark interactions and the restoration of chiral symmetry \cite{Detar:1995hk, Laermann:2003cv}.  
Conversely, in the hadronic phase, vector channel fluctuations are enhanced due to strong quark correlations and the presence of bound states. Furthermore, lattice QCD studies suggest that in the QGP phase, quark number and electric charge are primarily carried by quark-like quasi-particles, shedding light on the reduction of fluctuations in this regime \cite{Karsch:2005ps}.  
Lattice QCD has been instrumental in advancing our understanding of the QCD phase diagram, the equation of state, and screening lengths in the plasma phase \cite{Laermann:2003cv}. These studies provide critical insights into quark number susceptibilities and spectral densities, which are essential for interpreting experimental results from heavy-ion collisions.  
Moreover, the behavior of $\chi$ near the critical endpoint (CEP) of the QCD phase diagram aligns well with lattice QCD simulations, reinforcing its significance as a probe of critical phenomena \cite{He:2008yr}. This underscores the importance of $\chi$ in exploring the QCD phase structure. While $\chi$ can be determined from using either the dressed quark propagator or the spectral function, our focus here is on transport properties within finite-temperature field theory, where real-time dynamics play a crucial role. Consequently, we prefer to employ the spectral function approach for our analysis.

The quark number susceptibility essentially measures the response of the system's quark number density $n_q$ to variations in the quark chemical potential $\mu$:
\begin{equation}\label{susceptibility}
       \chi =\left. \frac{\partial n_q}{\partial \mu}\right|_{\mu=0}\,.
\end{equation}
In QCD, $\chi$ represents fluctuations in the conserved quark number and quantifies the degree to which the quark number deviates from its average value in thermal equilibrium. These fluctuations, arising from both thermal and quantum effects, provide valuable insights into the behavior of conserved charges such as baryon number and electric charge. They also help to illuminate the dynamics of the QGP and the transition between the hadronic and deconfined phases. A key reference paper is \cite{Gottlieb:1987ac}. Intuitively, one expects a rapid change in $\chi$ around deconfinement as a large number of quarks get liberated. 

We follow \cite{Kunihiro:1991qu} here. To account for the variation in quark number, the system is analyzed within the framework of the grand canonical ensemble. In this ensemble, the quark number density is described by the following expression:
\begin{equation}\label{number density}
    n_q = \frac{1}{V} \frac{\text{Tr}\left(N e^{-\beta(H-\mu N)} \right)}{Z},
\end{equation}
where $H$ represents the total energy of the system, encompassing the kinetic energy of quarks and gluons as well as the interaction energy between quarks. The chemical potential $\mu$ regulates the quark number and establishes the imbalance between quarks $q$ and antiquarks $\bar{q}$ within the system. The vector channel is linked to excitations in the system that correspond to a conserved vector current. This is typically associated with the quark number current
\begin{equation}
      J^\mu = \bar{\psi} \gamma^\mu \psi\,,
\end{equation}
where, as usual, $\gamma^\mu $ are the Dirac gamma matrices, and $\psi$ represents the quark field. The particle number operator, $ N $, is defined as $ N = \int dV \, J_0(t, \mathbf{x}) $, where $ J_0 $ is the temporal component of the conserved current. By utilizing Eqs.~(\ref{susceptibility}) and (\ref{number density}), one can derive the following expression for the susceptibility:
\begin{equation}
    \chi=\frac{\beta}{V} \sigma_N ^2,
\end{equation}
with $\sigma_N^2 = \langle N^2 \rangle - \langle N \rangle^2$ and $\langle N \rangle = n_q V$. This implies that large-scale fluctuations in $N$ are associated with a higher $\chi$. Near the phase transition, the sharp increase in $ \sigma_N^2 $ signifies critical phenomena. These fluctuations are inherent to equilibrium systems and originate from the random thermal motion of particles. The response function, such as $ \chi $, quantifies the system's reaction to small perturbations, such as slight changes in $ \mu $.
If $\langle N \rangle^2$ is negligible compared to $\langle N^2 \rangle$, the variance simplifies to:
\begin{equation}
\sigma_N^2 \approx \langle N^2 \rangle\,.
\end{equation}
This approximation holds under the following conditions:
\subsubsection*{Condition 1: Small Ensemble Average $\langle N \rangle$}
\begin{itemize}
    \item At low temperature or chemical potential ($\mu \to 0$), the small number of quarks leads to $\langle N \rangle \approx 0$, making $\langle N \rangle^2$ negligible.
    \item In symmetric systems (e.g., $q$ and $\bar{q}$ numbers), $\langle N \rangle = 0$, so $\langle N \rangle^2 = 0$.
\end{itemize}

\subsubsection*{Condition 2: Dominance of Fluctuations}
\begin{itemize}
    \item Close to the QCD critical point, large fluctuations dominate due to long-range correlations, making $\langle N^2 \rangle \gg \langle N \rangle^2$.
    \item In the QGP phase, deconfined quarks exhibit significant fluctuations, ensuring $\langle N^2 \rangle \gg \langle N \rangle^2$.
\end{itemize}

\subsubsection*{Condition 3: Statistical Properties}
For systems with Gaussian fluctuations around $\langle N \rangle \approx 0$, the variance is primarily determined by $\langle N^2 \rangle$, as $\langle N \rangle^2$ is negligible.
\\
\\
Therefore, under the condition that $\langle N \rangle^2$ is negligible, the final expression of $\chi$ simplified to
\begin{equation}
    \chi= \frac{\beta}{V} \langle N^2 \rangle_\beta\,.
\end{equation}
Now, it is known that
\begin{equation}
    \langle N^2 \rangle_\beta = \int dV \int dV' \langle J_0(t, \mathbf{x}) J_0(t, \mathbf{x}') \rangle_\beta\,.
\end{equation}
By translational invariance, the two-point correlation function $\langle J_0(t, \mathbf{x}) J_0(t, \mathbf{x}') \rangle_\beta$ depends only on the relative spatial separation $\mathbf{x} - \mathbf{x}'$. Therefore, one can set $\mathbf{x}' = 0$ and rewrite the expression:
\begin{equation}
    \langle N^2 \rangle_\beta = V \int dV \langle J_0(t, \mathbf{x}) J_0(t, \mathbf{0}) \rangle_\beta.
\end{equation}
Using this, the susceptibility finally takes the form
\begin{equation}
\chi = \beta \int dV \langle J_0(t, \mathbf{x}) J_0(t, \mathbf{0}) \rangle_\beta.
\end{equation}
Introducing the Fourier transform to analyze the correlation in terms of frequency and momentum:
\begin{equation}
J_0(t, \mathbf{x}) = \frac{1}{(2\pi)^4} \int d^3k \, d\omega \, e^{i(\mathbf{k} \cdot \mathbf{x} - \omega t)} \tilde{J}_0(\omega, \mathbf{k}),
\end{equation}
where $\tilde{J}_0(\omega, \mathbf{k})$ is the Fourier transform of $J_0(t, \mathbf{x})$. The correlation function $\langle J_0(t, \mathbf{x}) J_0(t', \mathbf{x}') \rangle_\beta$ in Fourier space is expressed as
\begin{equation}
\langle J_0(t, \mathbf{x}) J_0(t', \mathbf{x}') \rangle_\beta = \frac{1}{(2\pi)^4} \int d^3k \, d\omega \, S_{00}(\omega, \mathbf{k}) e^{i\mathbf{k} \cdot (\mathbf{x} - \mathbf{x}')} e^{-i\omega(t - t')}.
\end{equation}
Here, $S_{00}(\omega, \mathbf{k})$ is the spectral function that encodes the frequency and momentum dependence of the correlation. Using the temporal invariance, we can further set $t'=0$.  In the limit $\mathbf{k} \to 0$, the expression of $\chi$ reduces to
\begin{equation}\label{susceptibility-spectral function}
\chi = \beta \lim_{\mathbf{k} \to 0} \int \frac{d\omega}{2\pi} S_{00}(\omega, \mathbf{k}).
\end{equation}
It is well established that the Kubo-Martin-Schwinger (KMS) condition connects the symmetric correlator, \( S_{00}(\omega, \mathbf{k}) \) (the two-point function of currents), to the imaginary part of the retarded Green's function, \( G^R_{00}(\omega, \mathbf{k}) \),
\begin{equation}\label{spectral function-green function}
     S_{00}(\omega, \mathbf{k}) = -2 \, \text{Im} G_{00}^R(\omega, \mathbf{k}) \, \frac{1}{1 - e^{-\beta \omega}}.
\end{equation}
Here, \( (1 - e^{-\beta \omega})^{-1} \) originates from the thermal equilibrium population, and \( G^R_{00}(\omega, \mathbf{k}) \) represents the retarded Green's function, defined as
\begin{equation}
    G_{00}^R(\omega, k) = -i \int dt \int dV \, e^{i(\mathbf{k} \cdot \mathbf{x} - \omega t)} \, \theta(t) \, \langle \left[J_0(t, \mathbf{x}), J_0(0, \mathbf{0})]\right \rangle_\beta.
\end{equation}
From Eq.~(\ref{susceptibility-spectral function}), \( S_{00}(\omega, \mathbf{k}) \) can be expressed in terms of the inverse Fourier transform of \( \chi \). In the limit of long wavelengths and low frequencies, the expression simplifies to
\begin{equation}
    S_{00}(\omega, \mathbf{k})= \frac{2 \pi \delta(\omega) }{\beta} \chi\,.
\end{equation}
Using this relation, a connection between the imaginary part of the Green's function and the susceptibility can be derived from Eq.~(\ref{spectral function-green function}):
\begin{equation}\label{susceptibility-Im green func}
    \text{Im} G_{00}^R(\omega, \mathbf{k}) = -\frac{1 - e^{-\beta \omega}}{2}\frac{2 \pi \delta(\omega) }{\beta} \chi\,.
\end{equation}
The real and imaginary components of the Green's function are connected through the Kramers-Kronig relation, which arises as a fundamental consequence of causality, 
\begin{equation}\label{Kramers-Kronig}
    \text{Re} G_{00}^R(\omega, k) = \mathcal{P} \int \frac{d\omega'}{\pi} \frac{\text{Im} G_{00}^R(\omega', k)}{\omega' - \omega}.
\end{equation}
This connection emphasizes the interdependence of the system's dissipative (imaginary part) and reactive (real part) properties. At equilibrium, the lowest energy states (\( \omega \to 0 \)) dominate the susceptibility, and considering the long-wavelength limit, Eq.~(\ref{Kramers-Kronig}) simplifies to 
\begin{equation}
     \lim_{\mathbf{k} \to 0} \text{Re} G_{00}^R(0, \mathbf{k}) = \mathcal{P}\lim_{\mathbf{k}\to 0} \int \frac{d\omega'}{\pi} \frac{\text{Im} G_{00}^R(\omega', \mathbf{k})}{\omega'}.
\end{equation}
Using Eqs.~(\ref{susceptibility-spectral function}) and (\ref{spectral function-green function}), and expanding \( (1-e^{-\beta \omega'})^{-1} \) in the low-frequency limit, a direct relationship between the real part of the Green's function and the susceptibility can be obtained,  
\begin{equation}\label{susceptibility-Re green func}
    \chi=-\lim_{\mathbf{k}\to 0} \text{Re} G_{00}^R(0, \mathbf{k})\,.
\end{equation}
In later sections, we will compute $G_{00}^R(0, \mathbf{k})$ using holography and evaluate $\chi$ in the QGP medium.
%%%%%%%%%%%%%%%%%%%%%%%%%%%%%%%%%%%%%%%%%%%

 \subsection{ The Quark Number Susceptibility Calculation}
	From Eq.~(\ref{susceptibility-Re green func}), we focus on determining the $00$ component of the retarded Green's function, specifically with regards to a small, generic variation in momentum. 
For simplicity, we consider momentum along the $x_1$-axis, which matches the direction of the external magnetic field. By aligning momentum with the magnetic field, we are focusing on how the magnetic field affects the system's response. Since the quark number susceptibility is a scalar and does not care about direction, it does not matter which way the momentum points.\footnote{By taking the momentum in the $x_3$ direction, we get the following equation of motion for fluctuation: 
\begin{equation*}
	\partial_z^2 V_0 + \partial_z \left(\ln\left(\sqrt{-\mathcal{G}} f(z) G^{zz} G^{00}\right)\right)\partial_z V_0 - \frac{G^{33}}{G^{zz}} k^2 V_0 = 0\,.
 \label{susceptibilityEOM3}
\end{equation*}
This equation is similar to Eq.~(\ref{susceptibilityEOM}) except the last term. However, as shown in the Appendix \ref{appendixA}, both of these equations lead to the same expression for $\chi$. } This way, we can easily compute the susceptibility without losing sight of how the magnetic field influences the system. Eq.~(\ref{fluctuationEOM}), in radial gauge $V_z=0$, becomes
\begin{equation}
	\partial_z^2 V_0 + \partial_z \left(\ln\left(\sqrt{-\mathcal{G}} f(z) G^{zz} G^{00}\right)\right)\partial_z V_0 - \frac{G^{11}}{G^{zz}} k^2 V_0 = 0\,.
 \label{susceptibilityEOM}
\end{equation}

Our main focus is on understanding the behavior of susceptibility and diffusion coefficients at small momenta $k$. This can be done using the hydrodynamic expansions. In the hydrodynamics limit of large wavelength and small frequency, one arrives at the following expression of $\chi$ (see Appendix \ref{appendixA} for details):
\begin{eqnarray}
 \chi &=&  \left(\int _{z_h}^{0} \frac{dz}{\sqrt{-\mathcal{G}} f(z) G^{zz} G^{00}}\right)^{-1} \nonumber \\
 & = &  \left(\int _{0}^{z_h} \frac{dz}{ \frac{e^{-A(z)} \sqrt{e^{2 B^2 z^2+ 4 A(z)} L^4+ 4 B^2 \pi^2 z^4 {\alpha^\prime}^{2}} f(z)}{z}}\right)^{-1}\,.
\end{eqnarray}

\begin{figure}[htb!]
	\centering
	\includegraphics[height=7cm,width=11cm]{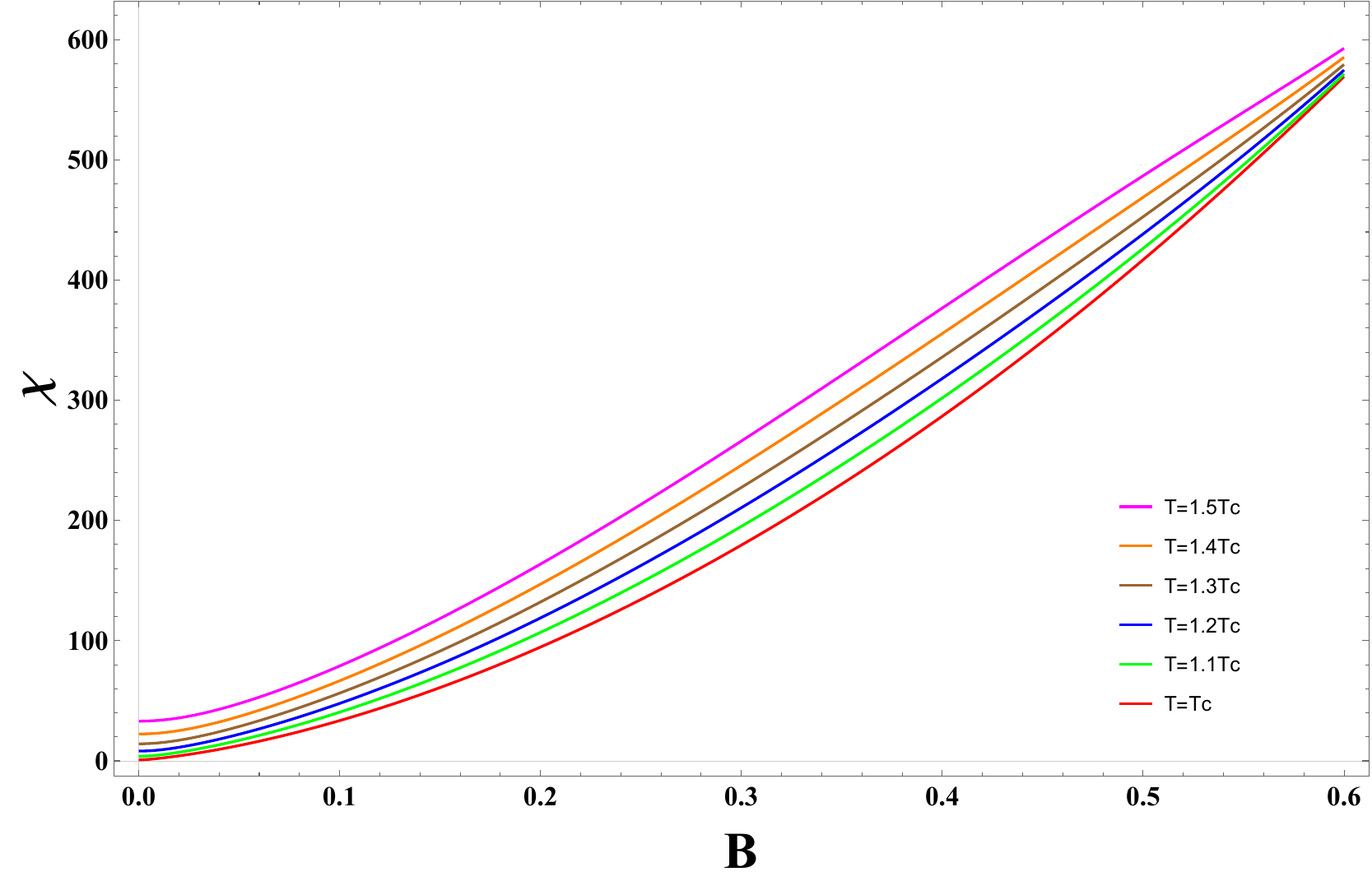}
	\caption{Quark number susceptibility $\chi$ as a function of magnetic fields.}
	\label{chivsB}	
\end{figure}
%%%%%%%%%%%%%%%%%%%%%%%%%%%%%%
The behavior of the susceptibility as a function of the magnetic field for different temperatures in the deconfined phase is shown in Figure~\ref{chivsB}. An increase in $\chi$ with $B$ is observed at all temperatures, which suggests that the quark number density becomes more sensitive to variations in the magnetic field strength. Our analysis further indicates a clear trend: as temperature increases, keeping the magnetic field constant, the quark number susceptibility also shows a corresponding increase. This trend strongly suggests that thermal influences play a significant role in the study of QGP.

%%%%%%%%%%%%%%%%%%%%%%%%%%%%%%
\begin{figure}[htb!]
	\centering
	\includegraphics[height=7cm,width=11cm]{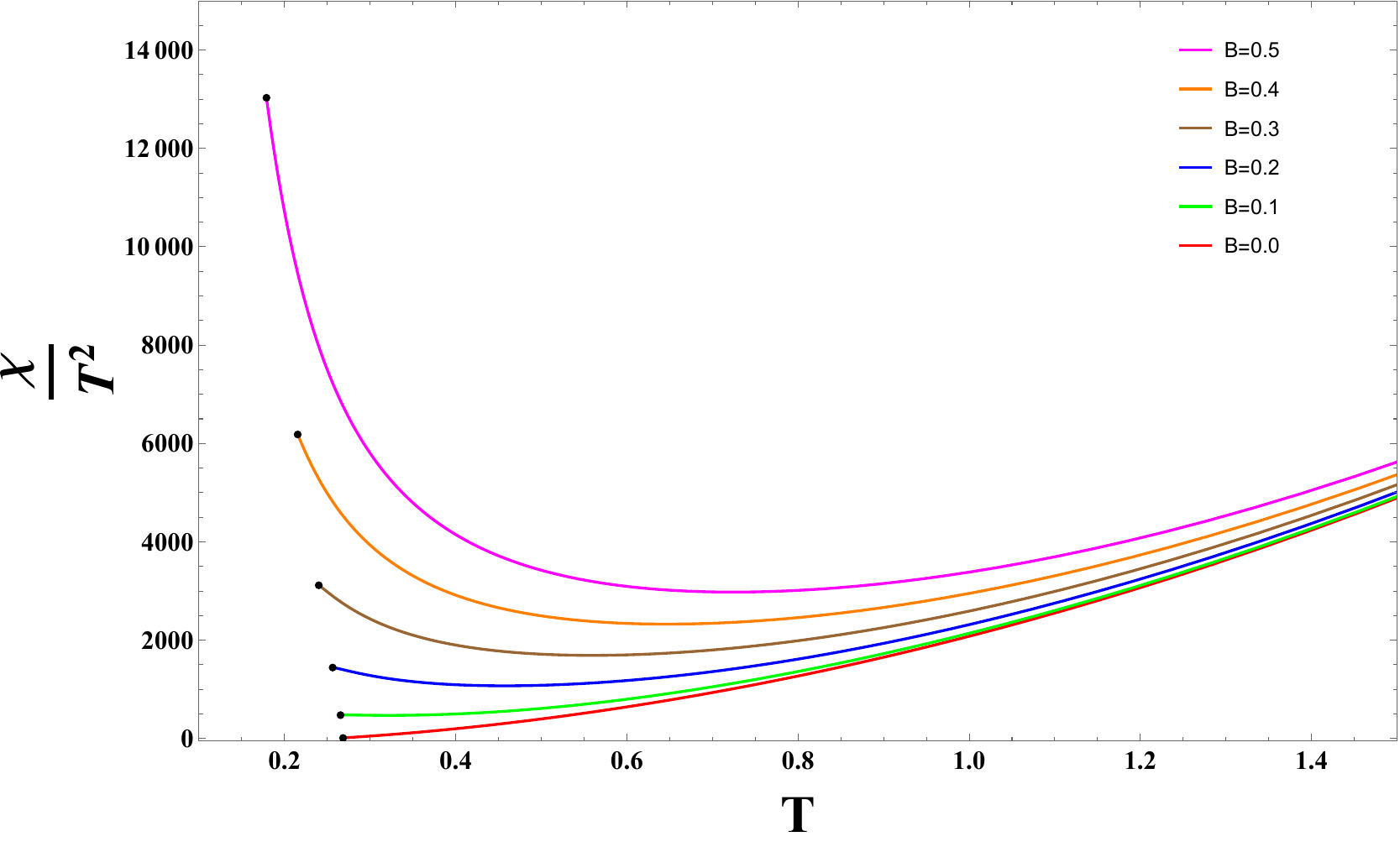}
	\caption{The rescaled quark number susceptibility $\chi/T^2$ as a function of temperature for different values of the magnetic field. In units GeV.}
	\label{chivsT}	
\end{figure}
%%%%%%%%%%%%%%%%%%%%%%%%%%%%%%
The behavior of the rescaled quark number susceptibility as a function of temperature for different magnetic field values is shown in Figure~\ref{chivsT}. Here the black dots denote the confinement-deconfinement transition temperature $T_{c}$. It is noteworthy that for sufficiently large temperatures, the behavior of the system becomes largely independent of the background magnetic field. This aligns well with theoretical expectations, as other scales are suppressed at large temperatures. Below the critical temperature ($T < T_{c}$), corresponding to the confined phase, the quark number susceptibility remains zero. This attribute stems from the confinement phase being dual to the thermal-AdS background, characterized by the absence of a horizon and temperature. This characteristic holds true across various large $N_c$ holographic QCD models, including our own. However, in the deconfinement phase (dual to a black hole), a significant increase in quark number susceptibility is observed, particularly in association with quarkonia pairs, notably in close proximity to the deconfinement temperature. Notably, a distinct peak in quark number susceptibility is evident near the deconfinement temperature. We further observe a distinct thermal behavior of $\chi$ at small and large magnetic field values, especially near the deconfinement temperature. In particular, for small $B$, $\chi$ first increases slightly with $T$ near the deconfinement temperature, whereas for large $B$, $\chi$ decreases substantially near the deconfinement temperature. Accordingly, our analysis suggests a distinct and potentially verifiable influence of $B$ on $\chi$ near the deconfinement temperature.

%%%%%%%%%%%%%%%%%%%%%%%%%%%%%%%%%%%%
\section{Diffusion Coefficient from  a Hydrodynamic Expansion}   
\label{sec4}
Now, we will formulate the analytical expression of the diffusion constant using hydrodynamic expansions.
Let us return to the equation of motion (\ref{parallelfluceq}). For fluctuation in the parallel direction, the equation of motion is given by
\begin{eqnarray}
    V'_1  + \frac{\gamma'}{\gamma} V'_1 - \frac{G^{tt}}{G^{zz}} \omega^2 V_1=0\,,
    \label{parallelEOM}
    \end{eqnarray}
where $\gamma$ is 
\begin{eqnarray}
    \gamma=\sqrt{-\mathcal{G}} f(z) G^{zz} G^{11}\,.
    \label{gammavalue}
\end{eqnarray}
We now consider the following  expansion of $V_1(z)$ (small $\omega$) near the horizon, 
\begin{eqnarray}
    V_1(z) = \biggl(1-\frac{z}{z_h}\biggr)^{-  \frac{ i \omega}{4 \pi T}}(F_0(z) + \omega F_\omega(z))\,.
    \label{hydrodynamicexpansiondiff}
\end{eqnarray}
The above expansion ensures the ingoing boundary condition for the mode $V_1$ at $z_h$. Expanding in powers of the variable $\omega$, the equation at the lowest order can be expressed as
\begin{eqnarray}
    \partial_z (\gamma F'_0)=0\,,
    \label{lowestordereqn}
\end{eqnarray}
implying that $\gamma F'_0$ must be a constant, i.e., $\gamma F'_0=C_1$. Accordingly, the general solution of $F_0$ is given by
\begin{eqnarray}
    F_0(z) = C_1 \int \frac{d z}{\gamma(z)}+C_2\,.
    \label{F0value}
\end{eqnarray}
The $F_0$ solution is singular at the horizon. Therefore, we must choose $C_1=0$ to avoid the singularity at $z=z_h$.
This leads to $F_0(z)= C_2$. After a first integration, the next to leading order order equation will be
\begin{eqnarray}
   \frac{i \omega}{4 \pi T }\left(1 - \frac{z}{z_h}\right)^ {-\frac{ i \omega}{4 \pi T} -1} \frac{C_2}{z_h} + \left(1 - \frac{z}{z_h}\right)^ {-\frac{ i \omega}{4 \pi T} } \omega F_\omega' = \frac{C_3}{\gamma}\,.
\end{eqnarray}
This leads to
\begin{eqnarray}
    F_\omega' = \frac{C_3}{\omega \gamma} \left(1 - \frac{z}{z_h}\right)^ {\frac{ i \omega}{4 \pi T} } -  \frac{i C_2}{4 \pi T (z_h - z)}\,.
\end{eqnarray}
The above equation can be further simplified by expanding the term $(1 - \frac{z}{z_h})^ {\frac{ i \omega}{4 \pi T} }$ for small $\omega$, i.e., $(1 - \frac{z}{z_h})^ {\frac{ i \omega}{4 \pi T} } = 1 + \mathcal{O} (\omega)$. Then the above equation leads to 
\begin{eqnarray}
    F_\omega = \int \frac{C_3}{\omega \gamma} dz + i \frac{C_2 z_h}{4 \pi T} \ln \left({1-\frac{z}{z_h}}\right)+C_4\,.
\end{eqnarray}
The first and second terms both again have a singularity at $z=z_h$. To avoid the singularity, one can choose
\begin{equation}
    \frac{C_3}{\omega \gamma}= \frac{i C_2}{4 \pi T (z_h - z)}\,,
\end{equation}
which allows us to find $C_3$ in terms of $C_2$. The determination of $C_4$, on the other hand, hinges on ensuring that $F_\omega (z=z_h)$ vanishes. However, its precise value will not be needed. Combining these results, the expression of $V_1(z)$ becomes
\begin{eqnarray}
    V_1(z)= C_2 \biggl(1-\frac{z}{z_h}\biggr)^{\frac{-i \omega}{4 \pi T}} \biggl(1 + \frac{8 \pi i \omega T}{z_{h}^2 L} \frac{ (L^4 + \Tilde{D}^2 z_{h}^4)}{e^{-3(B^2-a)z_{h}^2}}  \int_{z^*}^{z} \frac{dz}{\gamma(z)} 
    + \frac{i \omega z_h}{4 \pi T} \ln \left({1-\frac{z}{z_h}}\right)\biggr),
    \label{V1expression}
\end{eqnarray}
with
\begin{eqnarray}
    \Tilde{D} = 2  \pi \alpha' B e^{(2 a -B^2)z_{h}^2}\,,
\end{eqnarray}
and $z^*$ incorporates $C_4$ and is chosen ensuring $F_\omega (z=z_h)=0$. In our EBID model, to get the retarded Green's function we need $\frac{V_1 \partial_z V_1}{z^3}$ in the $z \rightarrow 0$ limit. Moreover, to get $\chi D$ we have to divide this expression by $\omega$ and take $\omega \rightarrow 0$ [see Eq.~(\ref{diffuexp})]. 

To compute the diffusion constant, we further require near boundary expansions of $V_1$ and $\partial_z V_1$. The near boundary expansion of $V_1$ is given by
\begin{eqnarray}
    V_1\big\rvert_{z\rightarrow 0} =C_2  \biggl(1 + \frac{8 \pi i \omega T}{z_{h}^2 L} e^{3(B^2-a)z_{h}^2} (L^4 + \Tilde{D} z_{h}^4) \int_{z^*}^{0} \frac{dz}{\gamma(z)} \biggr)\,,
    \label{V1z0exp}
\end{eqnarray}
which allows us to determine $C_2$ in terms of $V_1\big\rvert_{z\rightarrow 0}$. On the other hand, the derivative of $V_1$ becomes
\begin{align}
    \partial_z V_1 = & C_2 \frac{i \omega}{4 \pi T z_h} \left(1 - \frac{z}{z_h}\right)^ {-\frac{ i \omega}{4 \pi T} -1} \biggl(1 + \frac{8 \pi i \omega T}{z_{h}^2 L} e^{3(B^2-a)z_{h}^2} (L^4 + \Tilde{D}^2 z_{h}^4) \int_{z^*}^{z} \frac{dz}{\gamma(z)} \nonumber \\
    & + \frac{i \omega z_h}{4 \pi T} \ln \left({1-\frac{z}{z_h}}\right)\biggr) +  C_2 \biggl(1-\frac{z}{z_h}\biggr)^{-  \frac{ i \omega}{4 \pi T}} \biggl(\frac{8 \pi i \omega T}{z_{h}^2 L} e^{3(B^2-a)z_{h}^2} (L^4 + \Tilde{D}^2 z_{h}^4) \frac{1}{\gamma(z)} \nonumber \\
    & - \frac{i \omega}{4 \pi T} \frac{1}{1-\frac{z}{z_h}}\biggr)\,.
\end{align}
Dropping the term quadratic in $\omega$ and taking $z \rightarrow 0$, this reduces to 
\begin{eqnarray}
     \partial_z V_1\big\rvert_{z\rightarrow 0}= C_2 \biggl(\frac{8 \pi i \omega T}{z_{h}^2 L} e^{3(B^2-a)z_{h}^2} (L^4 + \Tilde{D}^2 z_{h}^4) \frac{1}{\gamma(0)}\biggr) + \mathcal{O}(\omega^2)\,.
     \label{partialV1z0exp}
\end{eqnarray}
Using Eqs.~(\ref{V1z0exp}) and (\ref{partialV1z0exp}), and ignoring some prefactors, the analytical expression of diffusion constant in the parallel case can be written as 
\begin{align}
    \chi D^\parallel = &  \Im \lim_{\omega\to 0} \frac{1}{\omega} \lim_{z \to 0} \frac{V_1 \partial_z V_1}{z^3} \nonumber \\
    = & \frac{T}{z_{h}^2} e^{3(B^2-a)z_{h}^2} (1 + \Tilde{D}^2 z_{h}^4)\,, 
\end{align}
where we have used $\gamma(0) = L^3/z^3$ and set the AdS length scale $L$ to $1$. 

The computational process remains nearly identical for the transverse polarizations. The primary distinction lies in determining the value of $C_3$ in terms of $C_2$ to prevent a singularity at $z=z_h$. Doing that for the transverse case, we get
\begin{eqnarray}
    C_3 =\frac{8\pi i \omega T }{z_{h}^2} e^{(2 B^2 -3 a)z_{h}^2} C_2\,.
\end{eqnarray}
Following the same approach as in the parallel case and discarding some prefactors, the analytical expression for the diffusion constant in the perpendicular direction is found to be
\begin{eqnarray}
    \chi D^\perp =  \frac{T}{z_{h}^2} e^{(2B^2-3a)z_{h}^2}\,.
\end{eqnarray}
As expected, when $B=0$, the parallel and perpendicular diffusion constants become equal, i.e., $D^\parallel=D^\perp$.

%%%%%%%%%%%%%%%%%%%%%%%%%%%%%%
\begin{figure}[htb!]
	\centering
	\includegraphics[height=7cm,width=11cm]{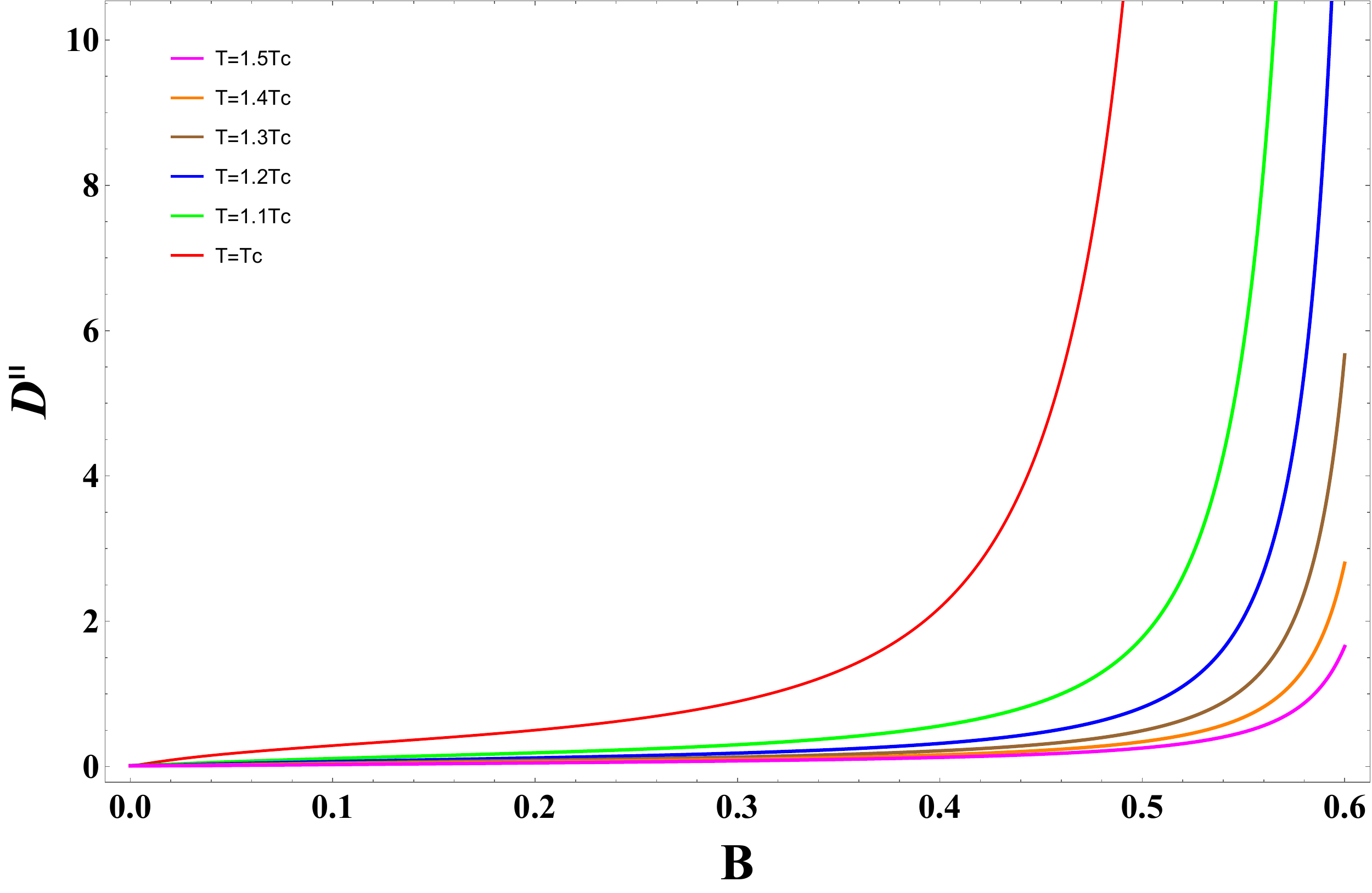}
	\caption{Heavy quark diffusion coefficient as a function of magnetic field for different temperatures in parallel directions to the magnetic field. In GeV units.}
	\label{DvsBparallel}	
\end{figure}
%%%%%%%%%%%%%%%%%%%%%%%%%%%%%%
%%%%%%%%%%%%%%%%%%%%%%%%%%%%%%
\begin{figure}[htb!]
	\centering
	\includegraphics[height=7cm,width=11cm]{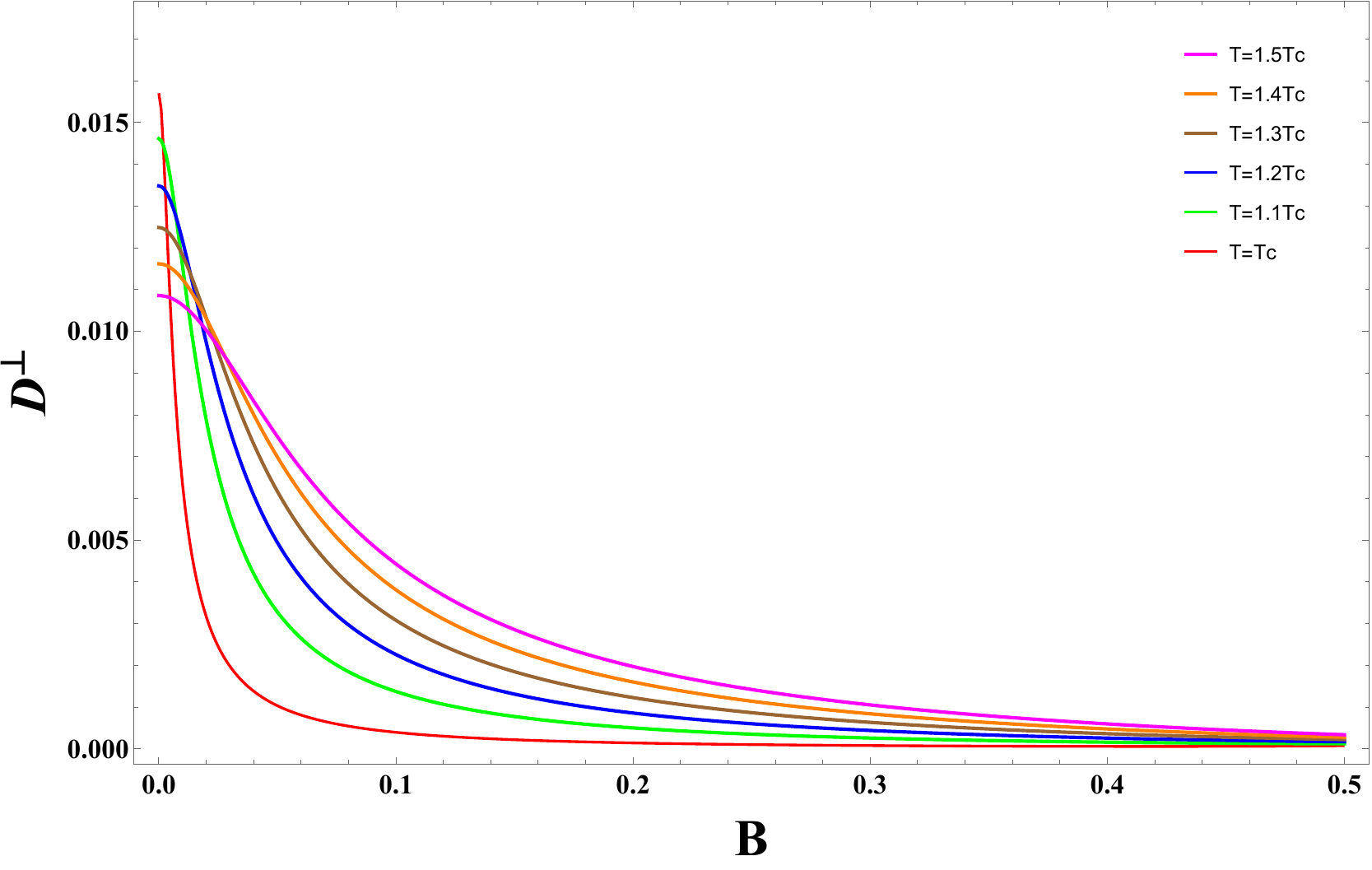}
	\caption{Heavy quark diffusion coefficient as a function of magnetic field for different temperatures in perpendicular directions to the magnetic field. In GeV units.}
	\label{DvsBperp}	
\end{figure}
%%%%%%%%%%%%%%%%%%%%%%%%%%%%%%

The effect of the magnetic field on the parallel diffusion coefficient $D^\parallel$ at different temperatures in the deconfined phase is shown in Figure~\ref{DvsBparallel}. It is observed that $D^\parallel$ increases with increasing magnetic field for all temperatures. This indicates that the presence of a magnetic field enhances the diffusion of heavy quarkonia in the parallel direction. The diffusion coefficient shows the most significant increase with the magnetic field at the deconfinement temperature $T=T_c$. As the temperature increases above $T_c$, the rate of increase of $D^\parallel$ with respect to the magnetic field $B$ decreases. This suggests that the diffusion of heavy quarkonia is more sensitive to the magnetic field at lower temperatures, particularly around $T_c$. At $T_c$, the rapid increase in $D^\parallel$ with $B$ might be due to the confinement-deconfinement phase transition that affects the quarkonia diffusion.

Figure~\ref{DvsBperp} represents the perpendicular heavy quark diffusion coefficient $D^\perp$ as a function of the magnetic field for different temperatures. All curves show a steep decline in $D^\perp$ as $B$ increases, indicating that the perpendicular diffusion coefficient decreases with increasing magnetic field strength. At high magnetic fields, $D^\perp$ values for all temperatures converge towards a similar value, indicating that the effect of temperature diminishes at high $B$. This behavior is consistent with the physical expectation that a perpendicular magnetic field suppresses the transverse motion of charged particles (like quarkonia) more effectively as the field strength increases.

When comparing this to the parallel case, we see that $D^\perp$ shows a decreasing trend with the magnetic field, whereas in the parallel case, $D^\parallel$ increases with the magnetic field strength. This highlights the anisotropic nature of diffusion under the influence of magnetic fields. The suppression of $D^\perp$ indicates that the magnetic field has a constraining effect on the transverse diffusion of quarkonia, reducing their mobility more effectively in the perpendicular orientation compared to the parallel orientation.

%%%%%%%%%%%%%%%%%%%%%%%%%%%%%%
\begin{figure}[htb!]
	\centering
	\includegraphics[height=7cm,width=11cm]{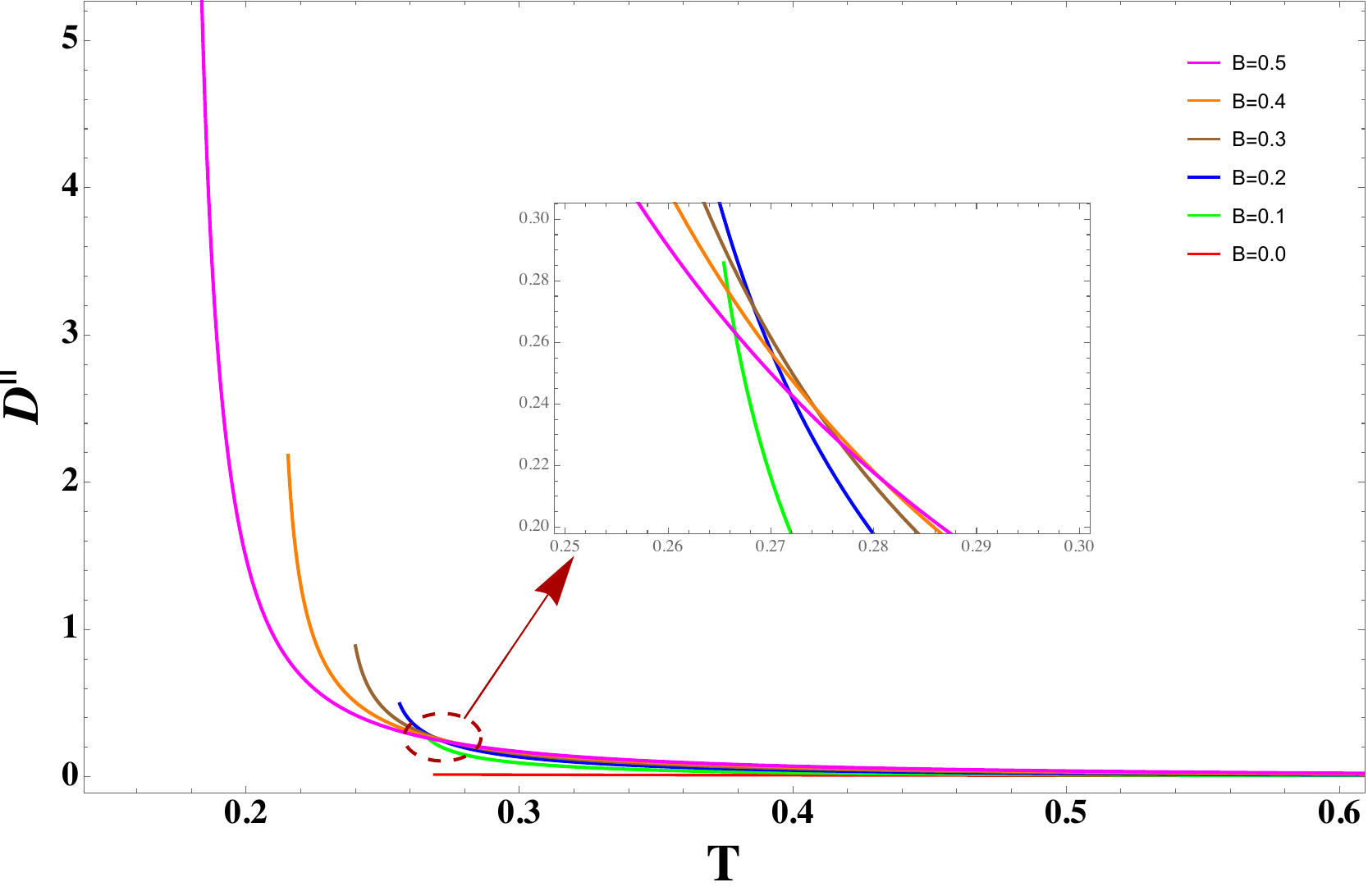}
	\caption{Heavy quark diffusion coefficient as a function of temperature for different values of the magnetic field in parallel directions to the magnetic field. In GeV units.}
	\label{DvsTparallel}	
\end{figure}
%%%%%%%%%%%%%%%%%%%%%%%%%%%%%%

Figure~\ref{DvsTparallel} shows the diffusion coefficient of heavy quarkonia in the parallel direction of the magnetic field as a function of temperature. Different colors represent different magnitudes of the magnetic field. We observe that $D^\parallel$ decreases as the temperature increases for all magnetic field values. The apparent crossover of the curves originates from the interplay between the two scales, temperature and magnetic field, and does not occur at a single fixed temperature. This feature is illustrated more clearly in the subplot of Figure~\ref{DvsTparallel}. At lower temperatures, the diffusion coefficient is higher, indicating that the heavy quarkonia have higher mobility in the medium. As the temperature increases, the diffusion coefficient rapidly decreases, suggesting that the medium becomes less conducive to the movement of heavy quarkonia. As the magnetic field strength increases the initial value of $D^\parallel$ at low temperatures increases. This suggests that stronger magnetic fields enhance the mobility of heavy quarkonia at lower temperatures. At higher temperatures, the differences between the diffusion coefficients for different magnetic field strengths become less pronounced. The magnetic field seems to enhance the diffusion coefficient at low temperatures but has a limited effect as the temperature increases.

%%%%%%%%%%%%%%%%%%%%%%%%%%%%%%
\begin{figure}[htb!]
\begin{minipage}[b]{0.49\linewidth}
\centering
\includegraphics[width=\linewidth]{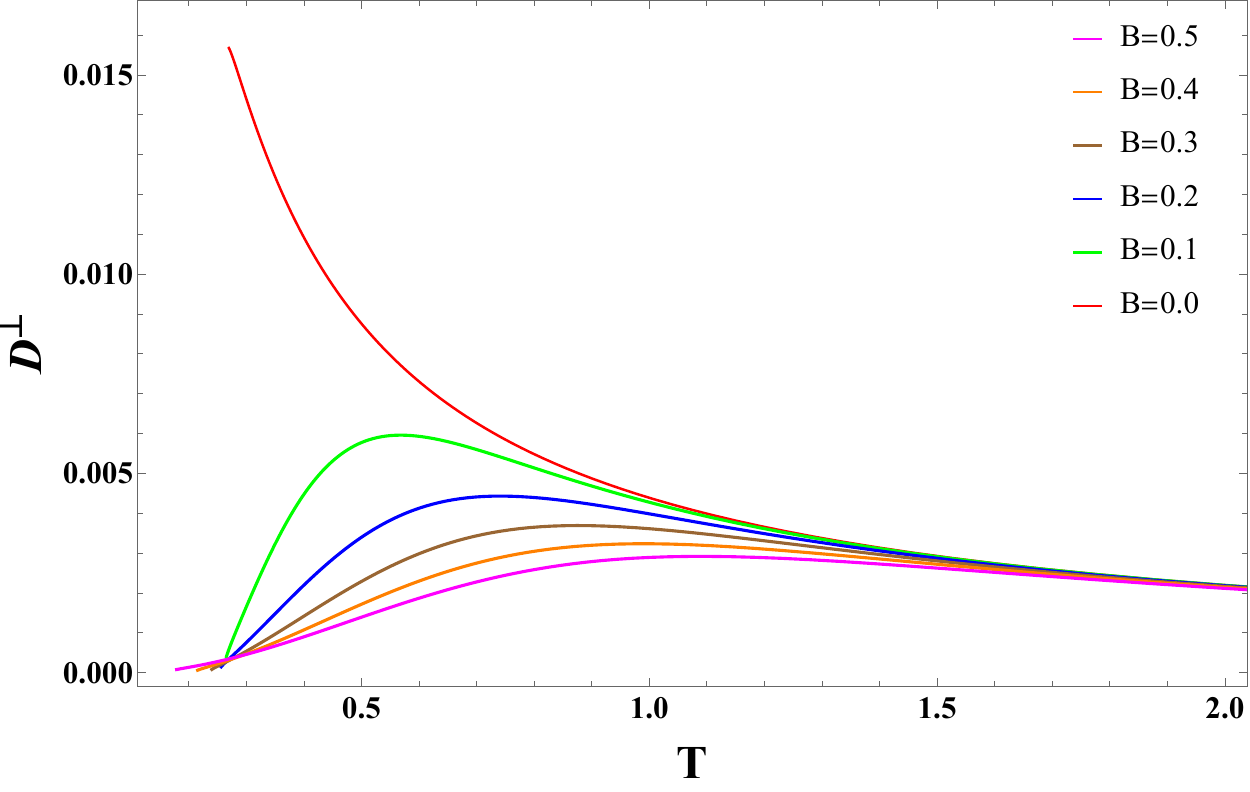}
\caption{\small Heavy quark diffusion coefficient as a function of temperature for different values of the magnetic field in perpendicular directions to the magnetic field. In $\text{GeV}$ units.}
\label{DvsTperp}
\end{minipage}
\hspace{0.2cm}
\begin{minipage}[b]{0.49\linewidth}
\centering
\includegraphics[width=\linewidth]{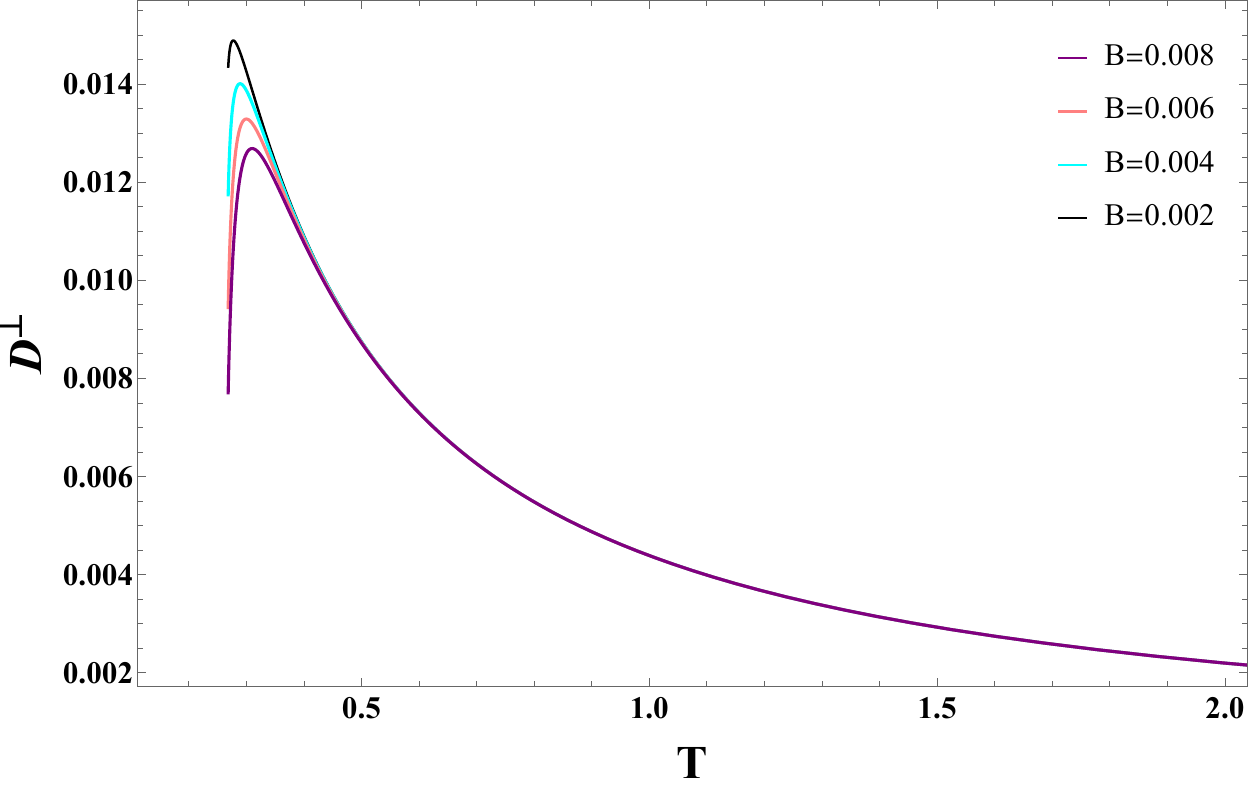}
\caption{\small Heavy quark diffusion coefficient as a function of temperature for lower values of the magnetic field in perpendicular directions to the magnetic field. In $\text{GeV}$ units.}
\label{DvsTperploerB}
\end{minipage}
\end{figure}
%%%%%%%%%%%%%%%%%%%%%%%%%%%%%%

Figure~\ref{DvsTperp} shows the diffusion coefficient of heavy quarkonia in the perpendicular direction of the magnetic field as a function of temperature. Here the same color coding is used as in Figure~\ref{DvsTparallel}. For a zero magnetic field, the diffusion coefficient starts at low temperatures and decreases monotonically with increasing temperature. For a nonzero magnetic field, the diffusion coefficient $D^\perp$ initially increases with temperature, reaches a peak and then decreases. The peak of the diffusion coefficient shifts to a higher temperature as the magnetic field increases. At low temperatures, stronger magnetic fields (higher $B$) result in a lower diffusion coefficient.
At intermediate temperatures, the diffusion coefficient peaks, and this peak is more pronounced and occurs at lower temperatures for weaker magnetic fields. At high temperatures, the diffusion coefficient values converge for different magnetic field strengths, showing minimal dependency on $B$. This transition from monotonic behavior to nonmonotonic behavior as the magnetic field is switched on is more clearly illustrated in Figure~\ref{DvsTperploerB}, where relatively lower values of the magnetic field are considered. Essentially, we observe a gradual transition in the presence of $B$.

Comparing the results with the parallel case, we observe that in the parallel case, stronger magnetic fields enhance the diffusion coefficient at low temperatures (near the deconfinement temperature) and have a diminishing effect at high temperatures. The monotonic decrease of $D^\parallel$ with temperature suggests that thermal agitation reduces the alignment and coherence of the quarkonia movement with the magnetic field, leading to decreased mobility which also suggests that quarkonia remain mostly in the diffusive regime, with interactions persisting even at higher temperatures. In the perpendicular case, stronger magnetic fields suppress the diffusion coefficient at low temperatures, and optimal diffusion occurs above the deconfinement temperatures for weaker fields. Therefore, in the perpendicular direction, the diffusion coefficient initially increases with temperature, indicating enhanced mobility near the deconfinement temperature, before decreasing at higher temperatures. The non-monotonic behavior of $D^\perp$ with temperature suggests a complex interplay between thermal energy and magnetic field effects. At low temperatures, the magnetic field restricts movement perpendicular to it, but as the temperature increases, thermal agitation initially enhances diffusion until a peak is reached. Beyond this peak, further increases in temperature disrupt coherence and reduce the diffusion coefficient. The perpendicular case shows a clear transition from a strongly diffusive regime at low temperatures through a more ballistic-like regime at intermediate temperatures (indicated by the peak in $D^\perp$) and back to a diffusive regime at higher temperatures due to increased thermal agitation.
 %%%%%%%%%%%%%%%%%%%%%%%%%%%%%%
\begin{figure}[htb!]
	\centering
 \subfigure[Parallel Case]{\includegraphics[width=0.42\linewidth]{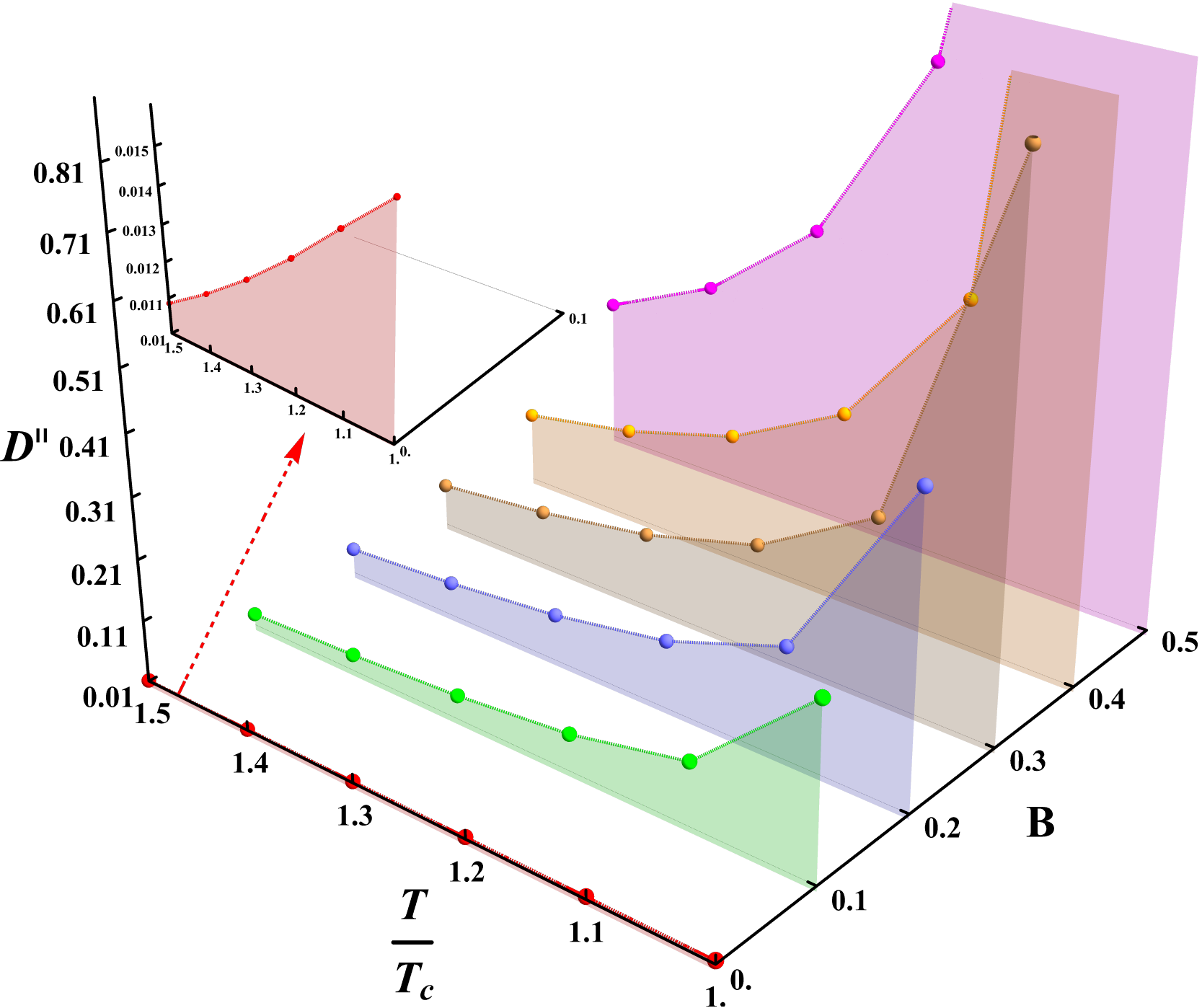}}
  \subfigure[Perpendicular Case]{\includegraphics[width=0.46\linewidth]{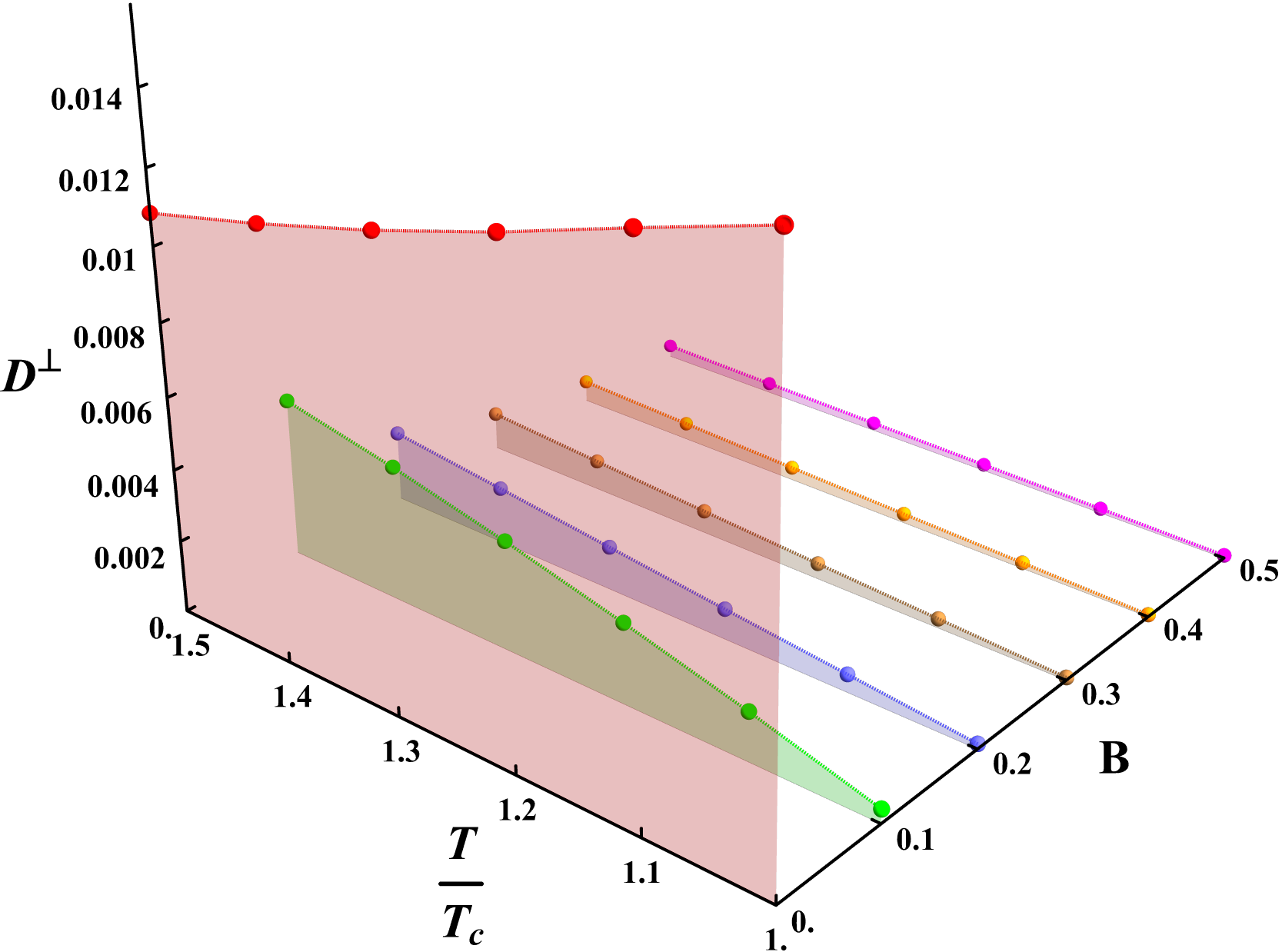}}
	\caption{\small Three-dimensional plot of diffusion coefficient as a function of temperature and magnetic field in the parallel and perpendicular directions. Red, green, blue, brown, orange, and magenta plots correspond to $B=0$, 0.1, 0.2, 0.3, 0.4, and  0.5, respectively. In units of $\text{GeV}$. }
	\label{3Dplot}	
\end{figure}
%%%%%%%%%%%%%%%%%%%%%%%%%%%%%%

To make the analysis more complete, in Fig.~\ref{3Dplot}, we have shown the three-dimensional plot illustrating how diffusion coefficients change with the magnetic field and temperature for parallel and perpendicular directions to the magnetic field. For the zero magnetic field, the diffusion coefficients are the same for both cases. However, at nonzero magnetic fields, we observe anisotropic behavior. Specifically, as the temperature increases, the diffusion coefficient decreases for the parallel magnetic field case but increases for the perpendicular case. Additionally, at a fixed temperature, the strength of the diffusion coefficient increases for the parallel case and decreases for the perpendicular case with the magnetic field. This behavior can be attributed to the way the magnetic field affects particle motion, enhancing diffusion in one direction while restricting it in another, leading to the observed anisotropy.

It would be interesting to compare the results of the diffusion constants in our self-consistent EBID model with those of the Born-Infield based soft wall model. In the soft wall holographic model \cite{Dudal:2018rki}, it was observed that the parallel diffusion coefficient $D^\parallel$ enhances with the magnetic field, while the transverse diffusion coefficient $D^\perp$ exhibits a strong suppression.  Moreover, at small temperatures, both $D^\parallel$ and $D^\perp$ tend to increase with temperature and then decrease for higher temperatures.
While the diffusion coefficients in the self-consistent EBID model share a qualitative agreement with the soft wall model predictions of  \cite{Dudal:2018rki} in terms of the magnetic field dependence, i.e., an enhancement of $D^\parallel$ and suppression of $D^\perp$ with increasing magnetic field, however, it provides a more subtle temperature dependence. In particular, $D^\parallel$ monotonically decreases with increasing temperature for all magnetic field values, indicating a stronger resistance to longitudinal motion at higher temperatures. On the other hand, similar to the soft wall model, $D^\perp$ displays a non-monotonic thermal behavior at finite magnetic fields; it initially increases with the temperature, reaches a maximum, and then decreases. 

These aforementioned results can be interpreted as granting further support for the underlying rationale of soft-wall model construction: in the absence of a fully self-consistent framework, qualitative insights into strongly coupled systems can still be reliably extracted from its soft-wall analog.

%%%%%%%%%%%%%%%%%%%%%%%%%%%%%%%%%%%%
\section{Diffusion Coefficient from a  Hanging String Approach}
\label{sec5}
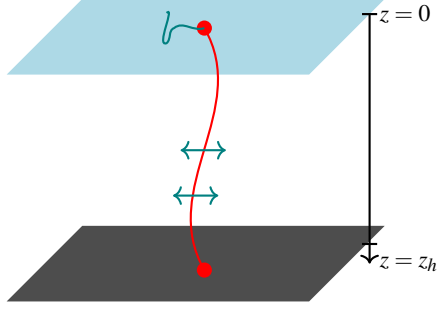
\begin{figure}[htb!]
    \centering
   \begin{tikzpicture}
        \definecolor{lightblue}{rgb}{0.68, 0.85, 0.90}
\definecolor{darkgray}{rgb}{0.3, 0.3, 0.3}

     \fill[lightblue] (-2,1) -- (2,1) -- (3,2) -- (-1,2) -- cycle;
    \fill[darkgray] (-2,-2) -- (2,-2) -- (3,-1) -- (-1,-1) -- cycle;
    \draw[thick,->] (2.8,1.8) -- (2.8,-1.5);
    \draw (2.8,1.8) node[right] {\footnotesize $z=0$};
    \draw (2.8,-1.5) node[right] {\footnotesize $z=z_h$};
    \draw[thick,-] (2.7,1.8) -- (2.9,1.8);
    \draw[thick,-] (2.7,-1.24) -- (2.9,-1.24);
     \draw[red,thick] (1,2,1) to[out=-60,in=120] (1,-1.2,1);
    \fill[red] (1,2,1) circle (0.1);
    \fill[red] (1,-1.2,1) circle (0.1);
     \draw[teal,thick,<->] (0.3,0) -- (0.9,0);
    \draw[teal,thick,<->] (0.2,-0.6) -- (0.8,-0.6);
    \coordinate (A) at (0.5,2.2,1);
    \coordinate (B) at (0.5,1.8,1);
\coordinate (C) at (0.6,2,1);
\coordinate (D) at (1,2,1);
\draw[teal,thick] (A) to[out=40,in=200](B) to[out=40,in=200] (C)to[out=40,in=200] (D);
   \end{tikzpicture}
    \caption{A schematic diagram of a long string stretches from the boundary to the horizon. Due to the Hawking-Unruh effect, transverse fluctuations arise along the string that gets propagated back toward the boundary, where they cause the endpoint of the string to undergo Brownian motion.}
    \label{hangingstring}
\end{figure}

Now, we will explore the diffusion coefficients of heavy quarks in QGP using the hanging string approach. In this method, a heavy quark in a stationary state is modeled as an on-shell external quark, represented by a string anchored at the boundary and hanging toward the horizon; a schematic diagram is presented in Fig.~\ref{hangingstring}. Conversely, a lighter quark, which is off-shell, is depicted as an open string initially stretched between the AdS boundary and a point slightly above the horizon. For both cases, thermal fluctuations induced by the black hole cause irregular motion at the endpoints on the boundary. The motion of the quark is determined by the dynamics of the string worldsheet. By examining the string's fluctuations in the background geometry, we can study Brownian motion and important transport coefficients, such as the diffusion constant. This helps us to understand how heavy quarks lose energy in strongly coupled plasma. Extensive research has been undertaken to quantify this energy loss using methods that integrate perturbative QCD with hydrodynamic approaches. Comprehensive reviews of these methods are provided in \cite{Majumder:2007iu, Casalderrey-Solana:2007xex}. Significant progress has also been made in modeling quark energy loss using holography. Initial studies on heavy quark diffusion in \(\mathcal{N}=4\) supersymmetric Yang-Mills (SYM) were outlined in \cite{Casalderrey-Solana:2006fio}, followed by several other foundational works \cite{Herzog:2006gh, Gubser:2006bz, Friess:2006fk, Liu:2006nn, Chernicoff:2006hi, Dudal:2018rki,deBoer:2008gu, Fischler:2012ff}.  The initial formulation for using strings with endpoints anchored both at the boundary and the horizon as a method for investigating string dynamics was introduced in \cite{deBoer:2008gu, Fischler:2012ff}. Over the years, various studies have explored the quark dynamics in different holographic settings \cite{Banerjee:2013rca, Chakrabortty:2016xcb, Atmaja:2010uu, Chernicoff:2009ff}.

To describe the dynamics of the string stretching radially from the boundary, positioned at $z=\epsilon$, down to the black hole horizon at $z=z_h$ \cite{Fischler:2012ff}, we employ the Nambu-Goto action:
\begin{eqnarray}
    S_{NG}=\frac{1}{2 \pi \alpha'} \int d \tau d \sigma \sqrt{-\gamma} \,,
\end{eqnarray}
where $\alpha'$ denotes the string tension, $\gamma=\det(\gamma_{ab})$, and $\gamma_{ab}=(g_s)_{mn} \partial_a X^m \partial_b X^n $ represents the induced metric on the string worldsheet. Here, $(g_s)_{mn}$ corresponds to the spacetime metric in the string frame. We can obtain the string frame metric from the Einstein frame metric via the transformation $(g_s)_{mn}=e^{\sqrt{2/3}\phi}g_{mn}$ \cite{Dudal:2018ztm, Dudal:2017max}. Utilizing this expression, the corresponding string frame metric is
\begin{eqnarray}
	& & ds^2=\frac{L^2 e^{2A_s(z)}}{z^2}\biggl[-g(z)dt^2 + \frac{dz^2}{g(z)} + dx_{1}^2+ e^{B^2 z^2} \biggl( dx_{2}^2 + dx_{3}^2 \biggr) \biggr]\,,
	\label{stringmetric}
\end{eqnarray}
where $A_s(z)=A(z)+\sqrt{\frac{1}{6}}\phi(z)$, with $A(z)$ and $\phi(z)$ as given in Eqs.~(\ref{asol}) and (\ref{phisol}).
For simplicity, we fix a static gauge, setting $\tau=t$, $\sigma=z$, and $X=X(\tau,\sigma)$. Up to quadratic order in the perturbations, the action simplifies and can be expressed as
\begin{eqnarray}
    S= \frac{1}{2 \pi \alpha'} \int dt dz \left[\sqrt{-g_{tt}g_{zz}}+ \sqrt{-\frac{g_{tt}}{g_{zz}}}\frac{g_{ii}}{2} X_{i}^{'2} -\sqrt{-\frac{g_{zz}}{g_{tt}}}\frac{g_{ii}}{2} \Dot{X}_{i}^{2}\right]\,,
    \label{perturbedaction}
\end{eqnarray}
where $X_{i}^{'}=\partial_z X_{i}$ and $\Dot{X}_{i}=\partial_t X_{i}$. Note that the first term does not contribute to the equations of motion of $X_{i}$. Accordingly, this term can be omitted from the action, giving 
\begin{eqnarray}
     S= \frac{1}{4 \pi \alpha'}\int dt dz \sqrt{-g_{tt}g_{zz}}\left[\frac{g_{ii}}{g_{zz}}X_{i}^{'2}-\frac{g_{ii}}{g_{tt}}\Dot{X}_{i}^{2}\right]\,.
\end{eqnarray}
After substituting the expression of the metric component explicitly, the action takes the form
\begin{eqnarray}
     S= \frac{1}{4 \pi \alpha'} \int dt dz \frac{e^{2 A_s(z)}g(z)}{z^2}\left[X_{1}^{'2}+e^{B^2z^2}\left(X_{2}^{'2}+X_{3}^{'2}\right)+\frac{1}{g^2(z)}\left(\Dot{X}_{1}^{2}+e^{B^2z^2}\left(\Dot{X}_{2}^{2}+\Dot{X}_{3}^{2}\right)\right)\right].
\end{eqnarray}
Since this action is quadratic in $X_{i}$, the corresponding equation of motion would be linear. It is convenient to express this action in terms of the tortoise coordinate $z_{*}$, defined by
\begin{eqnarray}
    dz_*= \frac{dz}{g(z)}\,.
    \label{tortoise coordinate}
\end{eqnarray}
In terms of $z_{*}$, the action takes the form
\begin{eqnarray}
     S= \frac{1}{4 \pi \alpha'} \int dt dz_{*} \frac{e^{2 A_s(z)}}{z^2}\left[\left(X_{1}^{'2}+e^{B^2z^2}\left(X_{2}^{'2}+X_{3}^{'2}\right)\right)+\left(\Dot{X}_{1}^{2}+e^{B^2z^2}\left(\Dot{X}_{2}^{2}+\Dot{X}_{3}^{2}\right)\right)\right],
     \label{tortoise}
\end{eqnarray}
where $X_{i}^{'}=\partial_{z_{*}} X_{i}$ and $\Dot{X}_{i}$ carries the usual meaning. The equation of motion of $X_i$ takes the form 
% (see Appendix~(\ref{KG}) for more details)
\begin{eqnarray}
    \left(\partial_{t}^2-\partial_{z_*}^2\right) X_i=0\,,
    \label{Klein-Gordon}
\end{eqnarray}
which is equivalent to the equation of motion of a massless Klein-Gordon field in flat spacetime.
Its two independent solutions are
\begin{eqnarray}
    X_{i}^{(in)}(z) = e^{-i \omega t} g_{i}^{(in)}(z) \sim e^{-i \omega (t+z_*)} \,,
    \label{xin} 
\end{eqnarray}
and
\begin{eqnarray}
    X_{i}^{(out)}(z) = e^{-i \omega t} g_{i}^{(out)}(z) \sim e^{-i \omega (t-z_*)}\,,
    \label{xout}
\end{eqnarray}
representing ingoing and outgoing waves, respectively.

Expanding $ 1/g(z)$ near the horizon up to the first order and integrating, we get the following approximate expression of $z_*$:
\begin{eqnarray}
    z_*(z,z_h)=\frac{\left(-1 + e^{(-3a + B^2)z_h^2} + (3a - B^2)z_h^2\right)\log(z_h-z)}{2(3a - B^2)^2 z_h^3}.
    \label{zstar}
\end{eqnarray}
Quantum field quantization in curved spacetime typically results in a mode expansion of the form \cite{Birrell:1982ix}
\begin{eqnarray}
  X_i(t,z)= \sum_{\omega} \left[a_\omega l_\omega(t,z)+a_{\omega}^{\dagger}l_{\omega}^{*}(t,z)\right]  \,.
  \label{modeexpansion}
 \end{eqnarray}
The basis function $l_\omega(t,z)$ represents positive-frequency modes, which can be described as a combination of ingoing and outgoing waves with freely adjustable coefficients, i.e.,
\begin{eqnarray}
    l_\omega(t,z)=\mathcal{K} \left[g^{out}(z)+\mathcal{M} g^{in}(z) \right]e^{-i \omega t} \,.
    \label{fomega}
\end{eqnarray}
The constant $\mathcal{M}$ is determined by the mixed boundary condition at $z=\epsilon$, and it is generally found to be a pure phase, $\mathcal{M}=e^{i\phi}$ (see section \ref{subsec1}). Consequently, the amplitudes of the outgoing and ingoing modes are equal, which allows the black hole emitting Hawking radiation to maintain thermal equilibrium \cite{Hemming:2000as}. Meanwhile, the constant $\mathcal{K}$ is established by normalizing the modes using the standard Klein-Gordon inner product.

The Klein-Gordon inner product is defined for any functions $p_i(t,z)$ and $q_j(t,z)$ that satisfy the equations of motion as
\begin{eqnarray}
    (p_i,q_j)_{\sigma} = -\frac{i}{2 \pi \alpha'} \int_{\sigma} \sqrt{h} n^\mu g_{ij} \left(p_i \partial_\mu q_{j}^*-\partial_\mu p_i q_{j}^*\right) \,,
    \label{KGinnerproduct}
\end{eqnarray}
where $\sigma$ represents a Cauchy surface in the $(t,z)$ section of the spacetime, $h$ is the induced metric on $\sigma$, and $n^\mu$ is the future-directed unit normal vector to $\sigma$. This definition ensures that the Klein-Gordon inner product does not depend on the specific choice of $\sigma$ \cite{Birrell:1982ix}. For simplicity, $\sigma$ is often taken to be a constant-$t$ slice with a unit normal vector field being $n^t=1/\sqrt{-g_{tt}}$ \cite{Giataganas:2018ekx}.

To determine the constant $\mathcal{K}$, the normalized Klein-Gordon inner product can be utilized,
\begin{eqnarray}
    (l_\omega,l_\omega) = -\frac{i}{2 \pi \alpha'} \int_{\sigma} \sqrt{\frac{g_{zz}}{-g_{tt}}} g_{ii} \left(l_\omega(z,t) \partial_t l_{\omega}^*(z,t)-\partial_t l_{\omega}(z,t)l_{\omega}^*(z,t)\right)  \,,
    \label{normalization}
\end{eqnarray}
where, as usual, $i=1$ corresponds to the string aligning parallel to the magnetic field and $i=2$ (or $3$) corresponds to the string aligning perpendicular to the magnetic field. The integral is primarily influenced by contributions from the near horizon region \cite{Caldeira:2020rir}.

For the parallel direction, we have
\begin{eqnarray}
    (l_\omega,l_\omega) = \frac{\omega}{ \pi \alpha'} \int_{zh(1+\epsilon)} dz \frac{e^{2 A_s(z)}}{z^2 g(z)} |l_\omega|^2 =1 \,.
    \label{normparallel}
\end{eqnarray}
Similarly, for the perpendicular direction, we have
\begin{eqnarray}
    (l_\omega,l_\omega) = \frac{\omega}{ \pi \alpha'} \int_{zh(1+\epsilon)} dz \frac{e^{2 A_s(z)+B^2 z^2}}{z^2 g(z)} |l_\omega|^2 =1\,.
    \label{normperp}
\end{eqnarray}
By expanding the integrand near the horizon, setting $z_h$ as $z_h(1+\epsilon)$\footnote{Here, $\epsilon$ is a small real number quantifying displacement away from the horizon.}, and considering the contribution from the leading-order term in $\epsilon$, the normalization constant is determined as
\begin{eqnarray}
    \mathcal{K}_\parallel = \sqrt{\frac{e^{2 a z_{h}^2-\sqrt{\frac{2}{3}}\phi(z_h)} \pi \alpha'}{4 \omega x \log(\frac{1}{\epsilon})}} \,, \hspace{10mm} \mathcal{K}_\perp = \sqrt{\frac{e^{(2 a-B^2) z_{h}^2-\sqrt{\frac{2}{3}}\phi(z_h)} \pi \alpha'}{4 \omega x \log(\frac{1}{\epsilon})}} \,,
    \label{Avalue}
\end{eqnarray}
where $x=\frac{e^y-y-1}{2 y^2 z_h}$ and $y=(-3 a+B^2)z_{h}^2$ and $\phi(z_h)$ is the solution of the dilaton field [Eq.~(\ref{phisol})] evaluated at $z=z_h$.

In the semiclassical regime, the string modes are thermally excited due to the Hawking radiation emitted by the worldsheet horizon. Specifically, these modes obey the Bose-Einstein distribution:
\begin{eqnarray}
    \langle a_{\omega}^\dagger a_{\omega'} \rangle = \frac{2 \pi \delta(\omega-\omega')}{e^{\beta \omega}-1} \,.
    \label{BEcondition}
\end{eqnarray}
Using this result alongside the mode expansion provided in Eq.~(\ref{modeexpansion}), a general expression for the displacement squared of the Brownian particle can be derived. To make the theory well-defined near the horizon, we introduce a small cut-off near the horizon. Then imposing Neumann boundary condition at the point $z=z_h-\epsilon$, we get
\begin{equation}
    \mathcal{M}=-\frac{g^{(out)'}(z)}{g^{(in)'}(z)} \Big|_{z=z_h -\epsilon}
   = e^{2 i \omega z_*(z)} \Big|_{z=z_h -\epsilon} \,.
    \label{nb} 
\end{equation}
Using the near-horizon approximation for the tortoise coordinate $z_*$, this simplifies to 
\begin{equation}
    \mathcal{M} = \epsilon ^{\frac{i \omega}{2 \pi T} } = \epsilon ^{\frac{i \omega}{2 \pi} \beta } = e^{-\frac{i \omega}{2 \pi} \beta \log(1/\epsilon)} \,.
    \label{nb1}
\end{equation}
Here, $\beta=1/T$ is the inverse temperature of the black hole. This shows that the spectrum becomes discrete, and the discreteness is given by 
\begin{equation}
    \Delta \omega = \frac{4 \pi^2}{ \beta \log(1/\epsilon)}\,.
\end{equation}
This leads to
\begin{eqnarray}
    \sum_{\omega} \rightarrow \log(1/\epsilon) \frac{\beta}{4 \pi^2} \int_{0}^{\infty} d \omega\,.
    \label{sumtoint}
\end{eqnarray}
To proceed, we first identify the position of the heavy quark with endpoints of the string at the boundary, $z=z_c$, where $z_c$ is a UV cut-off near the boundary to make the mass of the heavy quark finite.
\begin{eqnarray}
 x_i(t)= X_i(t,z_c)&=& \sum_{\omega} \left[a_\omega l_\omega(t,z_c)+a_{\omega}^{\dagger}l_{\omega}^{*}(t,z_c)\right]  \, \nonumber \\
  &=&\log(1/\epsilon) \frac{\beta}{4 \pi^2}\int_{0}^{\infty} d \omega \left[a_\omega l_\omega(t,z_c)+a_{\omega}^{\dagger}l_{\omega}^{*}(t,z_c)\right]  \,. 
  \label{xiposition}
\end{eqnarray}
Here, we have used Eq.~(\ref{sumtoint}) to express the summation as an integral.
Then, it follows that
\begin{eqnarray}
    \langle x_i(t) x_i(0) \rangle = \log(1/\epsilon) \frac{\beta}{4 \pi^2}\int_{0}^{\infty} d\omega d\omega'\left[\langle a_\omega a_{\omega'}^\dagger \rangle l_\omega(t,z_c)l_{\omega'}(0,z_c)^{*}+\langle a_{\omega'}^\dagger a_\omega  \rangle l_{\omega}(t,z_c)^{*} l_{\omega'}(0,z_c)\right] \,.
    \label{average}
\end{eqnarray}
This expression exhibits an IR divergence arising from the zero-point energy, which persists even at zero temperature. To address this, we regularize the divergence by applying normal ordering, defined as $:a_\omega a_\omega^\dagger:=:a_\omega^{\dagger} a_\omega:$. After implementing this regularization, we obtain
\begin{eqnarray}
    \langle :x_i(t)x_i(0): \rangle &=& \log(1/\epsilon) \beta \int_{0}^{\infty}  \frac{d\omega}{4 \pi^2} \frac{1}{e^{\beta \omega}-1} \left[l_\omega(t,z_c) l_{\omega}(0,z_c)^*+l_{\omega}(t,z_c)^* l_\omega(0,z_c)\right] \,, \nonumber \\ 
    &=&\log(1/\epsilon) \beta \int_{0}^{\infty} \frac{d\omega}{4 \pi^2} \frac{2 |\mathcal{K}|^2 \cos(\omega t)}{e^{\beta \omega}-1} |g^{out}(z_c)+\mathcal{M} g^{in}(z_c)|^2 \nonumber \\ 
    & = & \langle :x_i(0)x_i(t): \rangle ^*
\end{eqnarray}
Similarly,
\begin{eqnarray}
    \langle :x_i(t)x_i(t): \rangle = \log(1/\epsilon) \beta\int_{0}^{\infty}  \frac{d\omega}{4 \pi^2} \frac{2 |\mathcal{K}|^2}{e^{\beta \omega}-1} |g^{out}(z_c)+\mathcal{M} g^{in}(z_c)|^2 =\langle :x_i(0)x_i(0): \rangle\,.
\end{eqnarray}
The regularized mean square displacement can be written as
\begin{eqnarray}
    s_{reg}^2(t) &=& \langle :\left[x_i(t)-x_i(0)\right]^2 : \rangle \nonumber\\
    &=& \langle :x_i(t)x_i(t): \rangle + \langle :x_i(0)x_i(0): \rangle -\langle :x_i(t)x_i(0): \rangle -  \langle :x_i(0)x_i(t): \rangle \,.
    \label{displacementsquare}
\end{eqnarray}
Then, one gets
\begin{eqnarray}
   s_{reg}^2(t) &=& \frac{2 \beta }{\pi^2} \int_{0}^{\infty} d\omega  \frac{\log(1/\epsilon)|\mathcal{K}|^2\sin^2(\frac{\omega t}{2})}{e^{\beta \omega}-1} |g^{out}(z_c)+\mathcal{M} g^{in}(z_c)|^2 \nonumber \\
   &=& \frac{2 \beta}{\pi^2} \int_{0}^{\infty} \frac{d\omega}{\omega}   \frac{|\mathcal{K}'|^2\sin^2(\frac{\omega t}{2})}{e^{\beta \omega}-1} |g^{out}(z_c)+\mathcal{M} g^{in}(z_c)|^2 \,.
   \label{displacementsq}
\end{eqnarray}
Here, we have used $|\mathcal{K}'|^2 = \log(1/\epsilon) |\omega \mathcal{K}|^2$. To proceed further, we first need to compute $\mathcal{M}$.

\subsection{Determination of $\mathcal{M}$ using mixed boundary conditions}
\label{subsec1}
When a magnetic field is included directly in the bulk metric, the choice of boundary condition depends on how the field influences the dynamics of a probe particle at the boundary. A Neumann boundary condition, which sets the radial derivative of the field to zero, implies no net flux of momentum in the radial direction and is suited for free Brownian motion; however, it neglects the directional influence of the magnetic field, making it unsuitable for capturing correlated fluctuations or Lorentz-like effects. A Dirichlet boundary condition, which fixes the field value at the boundary, overly constrains the particle, suppressing its interaction with the magnetic field and preventing it from responding dynamically. In contrast, a mixed boundary condition balances the fixed position and derivative constraints, allowing the particle to experience the Lorentz force induced by the bulk magnetic field while retaining the freedom to fluctuate in response to it. This makes the mixed boundary condition the most physically consistent choice, as it captures both the drag and Brownian motion effects, ensuring the boundary particle responds appropriately to the magnetic field's influence \cite{Herzog:2006gh}.

This imposes a boundary condition of the form
\begin{eqnarray}
    \Pi_i^z \big|_{z = z_c} = \frac{\partial \mathcal{L}}{\partial X_i'} = F_i \,,
\end{eqnarray}
where $\Pi_i^z$ represents the canonical momentum flux and $F_i=-F_{ij}\Dot{X}^j$ is the Lorentz force induced by the magnetic field \cite{Fischler:2012ff}.

In our scenario, the magnetic field is oriented along the $x_1$ direction. Consequently, the field's influence on motion is primarily observed in the transverse directions, i.e., in $(x_2, x_3)$ directions, while the longitudinal motion remains unaffected. This setup contrasts with our previous spectral function approach, where both $D_\parallel$ and $D_\perp$ were found to exhibit nontrivial dependencies. The boundary conditions then are 
\begin{eqnarray}
    \frac{2 e^{2 A_s(z)+B^2 z^2}}{z^2} g(z) X_2'=-B \Dot{X}_3 \,, \\
     \frac{2 e^{2 A_s(z)+B^2 z^2}}{z^2} g(z) X_3'= B \Dot{X}_2 \,,
     \label{boundaryconditions}
\end{eqnarray}
which can be diagonalized using $X_\pm = X_2 \pm i X_3$;
\begin{eqnarray}
    \frac{2 e^{2 A_s(z)+B^2 z^2}}{z^2} g(z) X_\pm'\mp i B \Dot{X}_\pm =0\,.
    \label{modeBCs}
\end{eqnarray}
Both $X_\pm$ fields comply with the mode decompositions described above and can be regarded as fundamental quantum fields for further analysis. Recall that $X_\pm$ can be represented as a combination of the outgoing and ingoing modes derived earlier, each with arbitrary coefficients. Adopting the convention established in Eq.~(\ref{fomega}), we express them as
\begin{eqnarray}
    X_\pm(t,z) = \mathcal{K}_\pm \left[g^{out}(z)+\mathcal{M}_\pm g^{in}(z) \right]e^{-i \omega t} \,.
    \label{xpm}
\end{eqnarray}
From Eq.~(\ref{modeBCs}), it follows that
\begin{eqnarray}
   \mathcal{M}_\pm = - \frac{\frac{2e^{2 A_s(z)+B^2 z^2}}{z^2} g(z) g^{(out)'}\mp B \omega g^{(out)}}{\frac{2 e^{2 A_s(z)+B^2 z^2}}{z^2} g(z)g^{(in)'}\mp B \omega g^{(in)}} \equiv e^{i \phi_\pm} \,.
   \label{Bbarvalue}
\end{eqnarray}
It is evident that $\mathcal{M}_\pm$ is a pure phase, as this follows directly from the relation $g^{(out)}=g^{(in)*}$. Keeping terms up to leading order in frequency, we find that
\begin{eqnarray}
    \mathcal{M}_\pm = \frac{2 e^{(-2 a + B^2)z_h^2+\sqrt{\frac{2}{3}}\phi(z_h)} \mp i B z_h^2}{2 e^{(-2 a + B^2)z_h^2+\sqrt{\frac{2}{3}}\phi(z_h)} \pm i B z_h^2}\,.
    \label{Bbarfinal}
\end{eqnarray}
The mean displacement squared is given by Eq.~(\ref{displacementsq}). Since the integral for $t \rightarrow \infty$ is primarily influenced by small $\omega$, it follows that
\begin{eqnarray}
    \int_{0}^{\infty} \frac{d\omega}{\omega}   \frac{\sin^2(\frac{\omega t}{2})}{e^{\beta \omega}-1} \rightarrow \frac{\pi t}{4 \beta} \hspace{5mm} \text{for} \hspace{2mm} t \rightarrow \infty \,.
    \label{integral}
\end{eqnarray}
Thus, up to a prefactor, the diffusion constant $D_\perp$ is essentially governed by the factor $|g^{out}(z_c)+\mathcal{M} g^{in}(z_c)|^2$. This explains why focusing solely on the low-frequency approximation of the mode solution was sufficient. In this low-frequency regime and substituting Eq.~(\ref{Bbarfinal}), we get
\begin{eqnarray}
    |g^{out}(z_c)+\mathcal{M} g^{in}(z_c)|^2 = \frac{16 e^{2[(-2a+B^2)z_h^2+\sqrt{\frac{2}{3}}\phi(z_h)]}}{4 e^{2[(-2a+B^2)z_h^2+\sqrt{\frac{2}{3}}\phi(z_h)]}+B^2 z_h^4} \,.
    \label{goutginmodsq}
\end{eqnarray}
Here we have used $g^{in/out} \approx 1$ for low-frequency limit. As we are calculating the transverse diffusion coefficient $D^\perp$ here, substituting Eqs.~(\ref{Avalue}), (\ref{integral}), and (\ref{goutginmodsq}) into Eq.~(\ref{displacementsq}) and simplifying, we get
\begin{eqnarray}
   s_{reg}^2(t) \simeq \frac{\alpha'}{x} \frac{2 e^{(2a+B^2)z_h^2+\sqrt{\frac{2}{3}}\phi(z_h)}}{4 e^{2 [B^2 z_h^2+\sqrt{\frac{2}{3}}\phi(z_h)]}+B^2 z_h^4 e^{4 a z_h^2}} t\,.
   \label{sreg}
\end{eqnarray}
As we know, $s_{reg}^2(t) = 2 D t$, then the expression gives the transverse diffusion coefficient,
\begin{eqnarray}
    D^\perp \simeq \frac{\alpha'}{ x} \frac{e^{(2a+B^2)z_h^2+\sqrt{\frac{2}{3}}\phi(z_h)}}{4 e^{2 [B^2 z_h^2+\sqrt{\frac{2}{3}}\phi(z_h)]}+B^2 z_h^4 e^{4 a z_h^2}} \,,
    \label{Dtransverse}
\end{eqnarray}
where $x=\frac{e^y-y-1}{2 y^2 z_h}$ and $y=(-3 a+B^2)z_{h}^2$ as defined earlier. 

%%%%%%%%%%%%%%%%%%%%%%%%%%%%%%
\begin{figure}[htb!]
\begin{minipage}[b]{0.5\linewidth}
\centering
\includegraphics[width=0.9\linewidth]{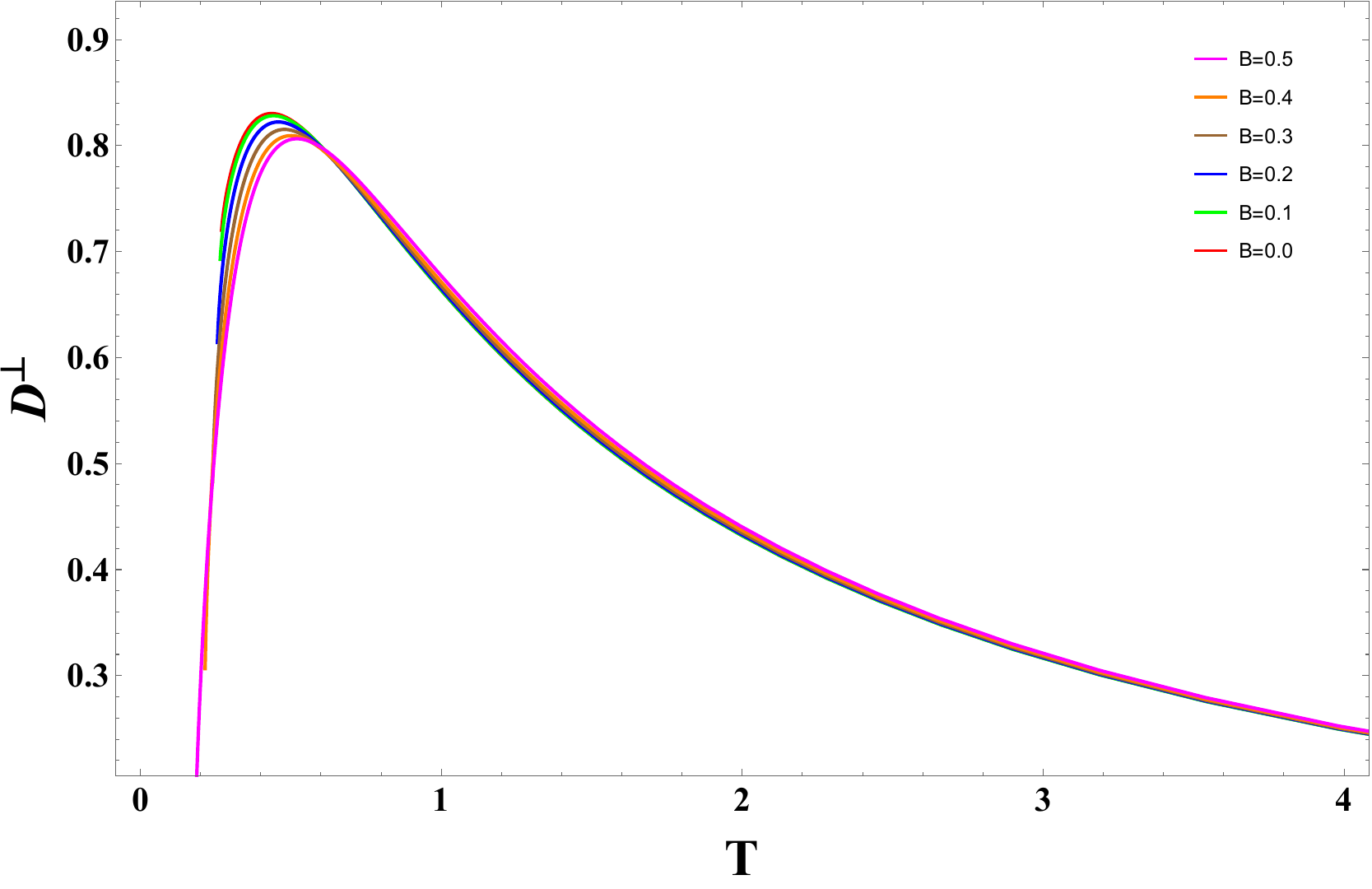}
\caption{\small Transverse diffusion coefficient as a function of temperature using the hanging string approach. Red, green, blue, brown, orange, and magenta plots correspond to  $B=0$, 0.1, 0.2, 0.3, 0.4, and  0.5, respectively. In units of $\text{GeV}$.}
\label{DperpvsTstring}
\end{minipage}
\hspace{0.2cm}
\begin{minipage}[b]{0.5\linewidth}
\centering
\includegraphics[width=0.9\linewidth]{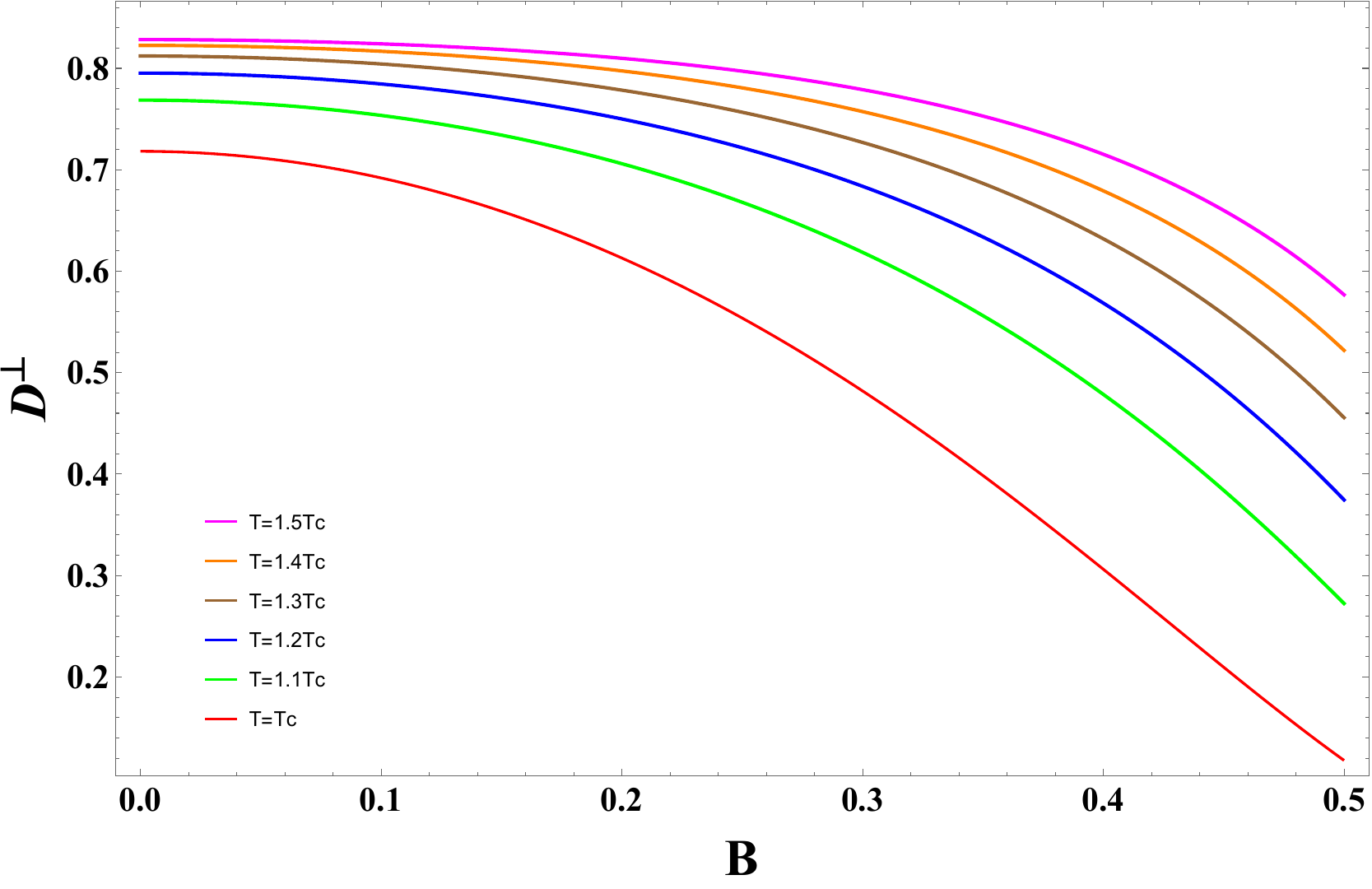}
\caption{\small Transverse diffusion coefficient as a function of magnetic field using the hanging string approach. Red, green, blue, brown, orange, and magenta plots correspond to  $T/T_c=1.0$, 1.1, 1.2, 1.3, 1.4, and  1.5, respectively. In units of $\text{GeV}$.}
\label{DperpvsBstring}
\end{minipage}
\end{figure}
%%%%%%%%%%%%%%%%%%%%%%%%%%%%%%

The behavior of the transverse diffusion coefficient, obtained by using the hanging string approach, as a function of temperature and magnetic field is illustrated in Figures~\ref{DperpvsTstring} and \ref{DperpvsBstring}, respectively. The results show that, at a fixed magnetic field, as the temperature increases, the diffusion coefficient increases and then decreases for higher temperatures. Additionally, at any fixed temperature, the diffusion coefficient decreases with the magnetic field. These results can be compared with the results obtained from the spectral function approach [Figures~\ref{DvsBperp} and \ref{DvsTperp}]. This means that higher temperatures enhance the mobility of the quark, while stronger magnetic fields constrain its motion, both of which are expected physical behaviors. Interestingly, this trend is quite similar to what we find using the spectral function approach, especially when the magnetic field is nonzero.

However, there is an important difference when we look at the case of a zero magnetic field, especially at low temperatures. In the spectral function approach, the transverse diffusion coefficient decreases with temperature when there is no magnetic field. But in the hanging string approach, it actually increases with temperature. This discrepancy likely comes from two main factors, i.e., the choice of the metric frame and the number of physical scales present in the system. The spectral function approach is usually carried out in the Einstein frame, while the hanging string approach is naturally formulated in the string frame as it requires the Nambu-Goto action. Since the string frame metric differs from the Einstein frame metric via the dilaton transformation $(g_s)_{mn}=e^{\sqrt{2/3}\phi}g_{mn}$, the effective geometry experienced by the string can vary significantly in the string frame. Maybe due to this, the two methods can capture the thermal effects in slightly different ways. Additionally, at zero magnetic field, temperature is the only relevant physical scale controlling the dynamics. Therefore, any temperature-dependent effects become more visible. In the hanging string approach, the string endpoint lies on the black hole region, making it more directly sensitive to thermal fluctuations. However, at finite magnetic field, the presence of an additional scale other than temperature tends to dominate the dynamics and can suppress these differences, especially at low temperatures, resulting in qualitatively similar behavior across both approaches. At high temperatures, as expected, both methods give the same qualitative behavior of the diffusion constant.

\section{Conclusion}
\label{sec6}
In this work, we considered a single-flavor holographic Einstein-Born-Infeld-dilaton (EBID) model to effectively analyze anisotropic imprints of the magnetic field on susceptibility and diffusion coefficients. In Einstein-Maxwell-dilaton (EMD) based holographic QCD models, Maxwell’s equations typically describe the behavior of electric and magnetic fields in a linear fashion. When considering quarkonium - bound states of quarks and antiquarks - it is important to note that while the constituent quarks carry electric charge, the overall quarkonium system is electrically neutral. As such, one might initially assume that external electromagnetic fields would have no direct influence on quarkonium. However, due to the internal structure of quarkonium, including the spatial distribution of its constituent charges - resembling an electric dipole - interactions with external fields can indeed occur. These interactions may result in nontrivial modifications to the properties or dynamics of the quarkonium state. To effectively capture such effects, a more advanced theoretical framework is necessary. The EBID action offers such a framework by incorporating nonlinear electromagnetic interactions and accounting for the internal structure of charged objects. In this context, the parameter $\ell_s\sqrt{q}$, where $q$ represents the quark or antiquark charge, serves as a smearing parameter that encodes information about the overall charge distribution. By extending the conventional Maxwell approach with EBI-like terms, the interaction between quarkonium and external electromagnetic fields can be modeled more accurately. In this work, based on the EBID action, we considered a self-consistent dynamical magnetized holographic QCD model of \cite{Jena:2024cqs}, to probe how a background magnetic field affects quark number susceptibility $\chi$ and diffusion coefficients $D$. For the record, we limited ourselves to diffusion along and perpendicular to the external magnetic field, to avoid getting all too cumbersome expressions.

To determine $\chi$, we used the spectral function approach combined with hydrodynamic expansion. Our results indicate that $\chi$ increases with both magnetic field and temperature, highlighting the important role these parameters play in the study of QGP. At high temperatures, $\chi$ becomes largely independent of the magnetic field, consistent with theoretical expectations. Our analysis further suggested a distinct thermal behavior of $\chi$ near the deconfinement temperature at small and large magnetic field values, thereby providing verifiable influence of $B$ on $\chi$.

The longitudinal $D^\parallel$ and transverse $D^\perp$ diffusion coefficients were then computed, and their closed-form expressions were obtained. We find that $D^\parallel$ increases with magnetic field at all temperatures, enhancing quarkonium mobility, with the most pronounced effects occurring near the critical temperature $T_c$. In contrast, $D^\perp$ decreases as magnetic field strength increases, revealing an anisotropic diffusion pattern in the presence of a magnetic field. Furthermore, $D^\parallel$ decreases with temperature, with the magnetic field's enhancing effect diminishing at high temperatures. $D^\perp$, on the other hand, shows nontrivial thermal behavior: it decreases monotonically with temperature at zero magnetic field, but under finite fields, it increases to a peak before declining, with the peak shifting to higher temperatures as $B$ increases. Essentially,  stronger magnetic fields suppress diffusion at low temperatures, maximize it at intermediate temperatures, and have little effect at high temperatures.

Finally, we also reconsidered the diffusion coefficient from a hanging string perspective. Here, the Hawking-Unruh effect causes transverse fluctuations along the string, which propagate toward the boundary, leading to Brownian motion at the endpoint. Since magnetic fields do not affect motion along their direction, as such we only computed the transverse diffusion coefficient. This differs from our previous spectral function approach, where we found that both longitudinal and transverse diffusion were influenced by the magnetic field. The thermal and magnetic field dependence of $D^\perp$ in this framework qualitatively agrees with the spectral function results at finite $B$. However, in the zero-field limit, $D^\perp$ decreases with temperature, contrary to the spectral function approach, where it increases, highlighting a difference between the two frameworks, for which we provided some intuition as to how we understand it.

\section*{Acknowledgments}
\label{sec7}
The work of S.~S.~J.~is supported by Grant No. 09/983(0045)/2019-EMR-I from CSIR-HRDG, India. The work of S.~M.~is supported by the core research grant from the Science and Engineering Research Board, a statutory body under the Department of Science and Technology, Government of India, under grant agreement number CRG/2023/007670.

\section*{Data Availability}
The data are not publicly available. The data are available from the authors upon reasonable request.

	\appendix
 \section{ Appendix }
 \label{appendixA}
In this Appendix, we provide the necessary details that go into the computation of the susceptibility. For this, we take the gauge field fluctuation equation of motion (\ref{susceptibilityEOM}) and rewrite it as
 \begin{equation}
    \partial_z(X_{tz} V_0') -k^2 p V_0=0\,,
    \label{EOMappendix}
\end{equation}
where
\begin{equation}
    X_{tz}=\sqrt{-\mathcal{G}} f(z) G^{zz} G^{00}\,,
\end{equation}
and $p= \sqrt{-\mathcal{G}} f(z)G^{00} G^{11}$ (or $p= \sqrt{-\mathcal{G}} f(z)G^{00} G^{33}$) when the momentum is chosen along the $x_1$ direction (or the $x_3$ direction).

Next, we consider the hydrodynamic expansion of the form,
 \begin{eqnarray}
	    V_0(z)= F_0(z) + k^2 F_{k^2}(z)+...\,.
     \label{hydrodynamicexpansionappendix}
	\end{eqnarray}

Substituting the value of $V_0(z)$ from (\ref{hydrodynamicexpansionappendix}) into (\ref{EOMappendix}), we get
 \begin{equation}
     (X_{tz}(F_0'(z)+k^2 F_{k^2}'(z)))' -k^2 p (F_0(z)+k^2 F_{k^2}(z))=0\,.
 \end{equation}
Equating the coefficients of various power of $k$, we get the following equation of motion for $F_0(z)$ and $F_{k^2}(z)$: 
\begin{equation}
    (X_{tz} F_0'(z))'=0\,.
    \label{F0EOM}
\end{equation}
\begin{equation}
	    (X_{tz} F_{k^2}'(z))' - p F_0(z) =0\,.
     \label{Fk^2EOM}
	\end{equation}
From Eq.~(\ref{F0EOM}), we get the following solution
\begin{equation}
    X_{tz} F_0'(z)=C_0\,.
    \label{condition}
\end{equation}
By integrating the above, we get the expression of $F_0(z)$ as
\begin{equation}
    F_0(z)= C_0 \int_{z_h}^{z} \frac{d z}{X_{tz}}\,,
    \label{F0sol}
\end{equation}
where $C_0$ is a constant of integration. Substituting (\ref{condition}) into (\ref{Fk^2EOM}), leads to
\begin{equation}
     (X_{tz} F_{k^2}'(z))' = p C_0 S(z)
\end{equation}
where $S(z)=\int_{z_h}^{z} \frac{ d\tilde{z}}{X_{tz}(\tilde{z})}$.
Substituting the above expressions into~(\ref{EOMappendix}) and comparing the coefficient of $k^2$:
\begin{align}
    k^2 p V_0 &= (X_{tz} V_0')' = k^2 (X_{tz} F_{k^2}'(z))' \nonumber \\
    &= k^2 p C_0 S(z)\,.
    \label{k2coefficient}
    \end{align}
This leads to
\begin{equation}
 V_0 = C_0 S(z)\,.
       \label{V0sol}
    \end{equation}  
 
Now, the quark number susceptibility is related to the retarded Green's function 
 \begin{eqnarray}
     \chi =  -\lim_{k,z \to 0} \Re [G_{tt}(\omega=0,k)]\,.
 \end{eqnarray}
The retarded Green's function is obtained by taking the second derivative of the on-shell boundary action with respect to the boundary value of the field. For our model, the boundary on-shell action is given by
\begin{eqnarray}
	S_{\text{on-shell,bdy}} & = & \lim\limits_{z\to 0}\int d^4x\frac{4}{z^3}V_0\partial_z V_0\,.
    \label{onshell}
\end{eqnarray} 
Putting the value of $V_0$ from (\ref{V0sol}) in the on-shell boundary action (\ref{onshell}), we get the retarded function as $G_{tt}(\omega=0,k)= -\frac{1}{S(0)}$. Then, using the explicit form of $X_{tz}$, we get 
\begin{eqnarray}
    \chi = \frac{1}{S(0)} =  \left(\int _{z_h}^{0} \frac{dz}{\sqrt{-\mathcal{G}} f(z) G^{zz} G^{00}}\right)^{-1}\,.
\end{eqnarray}
Here, we observe that $\chi$ is independent of $p$, which characterizes the direction of momentum. Consequently, the susceptibility remains the same for the momentum along both the $x_1$ and $x_3$ directions.

 \bibliography{mybib.bib}
\bibliographystyle{unsrt}
	
%\bibliographystyle{JHEP}
%\bibliography{mybib}
\end{document}